 \documentclass[final,1p,times]{elsarticle}

\usepackage{amssymb}
\usepackage{amsmath}
\usepackage{epstopdf}
\usepackage{url}
\usepackage{subcaption}
\usepackage[colorlinks=true,citecolor=blue,linkcolor=blue,urlcolor=blue]{hyperref}

\begin{document}


\begin{frontmatter}


\title{A Comprehensive Monte Carlo Simulation Tool on Electron Transport in Noble Gases and Liquids} 

\author[label1,label2]{Lei Cao}
\author[label2]{Guofu Cao}
\author[label3]{Yan Fan}
\author[label2]{Zhilong Hou}
\author[label1]{Yongsheng Huang\corref{corresponding}}
\author[label2]{Tao Liu}
\author[label2]{Fengjiao Luo}
\author[label3]{Hankun Ma}
\author[label2]{Xilei Sun\corref{corresponding}}
\author[label3]{Xiangming Sun}
\author[label2]{Jingbo Ye}
\author[label2]{Weixi Zhang}

\cortext[corresponding]{Corresponding author. sunxl@ihep.ac.cn (Xilei Sun), huangysh59@mail.sysu.edu.cn (Yongsheng Huang)}

\affiliation[label1]{organization={School of Science},
            addressline={Shenzhen Campus of Sun Yat-sen University},
            city={Shenzhen},
            postcode={518107},
            state={Guangdong},
            country={China}}
 
\affiliation[label2]{organization={Institute of High Energy Physics},
            addressline={Chinese Academy of Sciences},
            city={Beijing},
            postcode={100049},
            country={China}}

\affiliation[label3]{
            organization={College of Physical Science and Technology},
            addressline={Central China Normal University},
            city={Wu Han},
            postcode={430079},
            country={China}}
           

\begin{abstract}

 For the particle detectors based on noble gases or liquids, it is essential to understand the transport dynamic and the properties of the electrons. We report the development of a tool for electron transport in noble gases He, Ne, Ar, Kr, or Xe, and liquids Ar, Kr, or Xe. The simulation, implemented in C++ and MATLAB, is based on electron-atom collisions, including elastic scattering, excitation and ionization. We validate the program through assessing the electron’s swarm parameters, specifically the drift velocity and the diffusion coefficient. For electron transport in liquids, two models are discussed and both are used for the construction of the Monte Carlo framework based on the Cohen Leker theory. The results demonstrate the effectiveness and accuracy of the simulation tool, which offers a valuable support for detector design and data analysis.

\end{abstract}



\end{frontmatter}

\section{Introduction}
\label{Introduction}

The micro-physics study of electron transport in noble gases and liquids holds significance for particle detection technologies. The advancement of neutrino and dark matter detection relies chiefly on the use of liquid phase or gas-liquid dual phase detectors. Notable projects in this domain include XENONnT, which employs liquid xenon to probe the nature of dark matter\cite{XENONnT}, and LUX-ZEPLIN, an extension of the LUX experiment that continues to push the boundaries of direct dark matter detection using liquid xenon\cite{akerib2020lux}. Similarly, PandaX\cite{cao2014pandax} has contributed significantly to the field through its use of both gaseous and liquid xenon. MicroBooNE\cite{acciarri2017design}, SBND\cite{SBND} and ICARUS\cite{ICARUS} are three liquid argon neutrino detectors at Fermilab. Additionally, DUNE\cite{DUNE} and the DarkSide program\cite{DarkSide-PhysRevD.98.102006} have also adopted liquid argon as the detection medium for neutrino or dark matter detection. 

\indent Understanding electron drift and diffusion is important for the development of modern noble gas or liquid detectors. On the one hand, this work can be used to predict the electron properties that are poorly measured or not measured, e.g. the diffusion in LKr. On the other hand, using the parameters evaluated herein, simulation for the detectors including noble gases or liquids can be performed quickly.
 Though there have been some major advancements on the electron transport in noble gases and liquid Ar/Xe, there has been little information about the transport properties in LKr and the MC scattering models of noble liquids. The purpose of this paper is to provide a comprehensive Monte Carlo simulation for electron transport in noble gases He, Ne, Ar, Kr, Xe and liquids Ar, Kr, Xe. This work will be useful for future experiments, such as the proposed Circular Electron Positron Collider (CEPC)\cite{CEPC} and the detector on it, which has a TPC in its tracking system\cite{CEPC-TPC}.

\indent  In theoretical work, Cohen Leker et al.\cite{cohen1967theory} highlighted the differences in energy transfer and momentum transfer frequencies of hot-energy electrons in noble liquids during elastic scattering. The spectral function $S$ introduced in their work is the Fourier transform of the Van Hove space-time pair correlation function $g$ and is related to the particle momentum and energy for the structural effects in liquid and solid phases\cite{cohen1967theory}. This is referred as the CL theory. 
In addition, research by Wojcik and Tachiya\cite{Wojcik-M-Tachiya200320} focus on low-energy electron collisions in liquid argon. The interactions of the energy transfer and the momentum transfer are sampled based on the ratio of the energy transfer to the momentum transfer cross section. This is called the WT model in this paper.
Tattersall et al.\cite{tattersall2015monte} extended the WT model and proposed the SSMC model (Static Structure Monte Carlo) to optimize the electron collision in the energy region where the momentum transfer frequency exceeds the energy transfer frequency by sampling the energy transfer, the momentum transfer, and single-particle collisions (both transfer). These advancements in simulation models aim to enhance the accuracy of electron transport simulations in noble liquids.

\indent Detector simulation software is often used for gas and solid phase, such as Garfield++\cite{garfieldpp}, MagBoltz\cite{MagBoltz} and PyBoltz\cite{PyBoltz}. MagBoltz is a Monte Carlo simulation program written in FORTRAN for electron transport in the gas and gas mixtures. And PyBoltz is a Python-based adaptation of the original MagBoltz program. Efforts have been made to develop a simulation tool for electron transport in liquid Ar and Xe. Z. Beever et al.\cite{beever2024translate} conducted simulation studies on gaseous and liquid argon based on the WT model, and developed the simulation tool named TRANSLATE (TRANSport in Liquid Argon of near-Thermal Electrons). 
Yijun Xie et al.\cite{xie2024development} conducted Monte Carlo simulations of electron transport in liquid argon and liquid xenon in Geant4, which achieved a good agreement between the simulation and the experiment for electric fields ranging from 200 V/cm to 2000 V/cm in liquid argon and below 300 V/cm in liquid xenon.

\indent Despite these advancements, the Monte Carlo simulation of electron transport in noble liquids remains incomplete. This work aims to establish a comprehensive Monte Carlo tool for noble gases and liquids in the electric field from 10 V/cm to 2000 V/cm. By validating the key electron swarm parameters: drift velocity and diffusion coefficients, this research provides a reliable tool for simulating electron transport in noble gas helium, neon, argon, krypton, xenon and liquid argon, krypton, xenon, thereby offering strong support for the electron drift study in detectors and advancing research in the related fields.

\indent 
We begin this study with a brief introduction to the electron transport theory and the Monte Carlo simulation tools that are available currently. In Sec.\ref{section 2}, we analyze the Monte Carlo simulation framework for gas phase and transition that for the liquid phase by examining the disparities in electron transport between these two phases. In Sec.\ref{section 3}, we present the simulation results for the drift velocity and the diffusion coefficient, namely the electron swarm parameters in gases and liquids. The conclusion and future prospects are in Sec.\ref{section 4}.

\section{Principle and Simulation Framework}
\label{section 2}
\subsection{Gas phase simulation framework}

\begin{figure}[!t]
\centering
\includegraphics[width=0.9\textwidth]{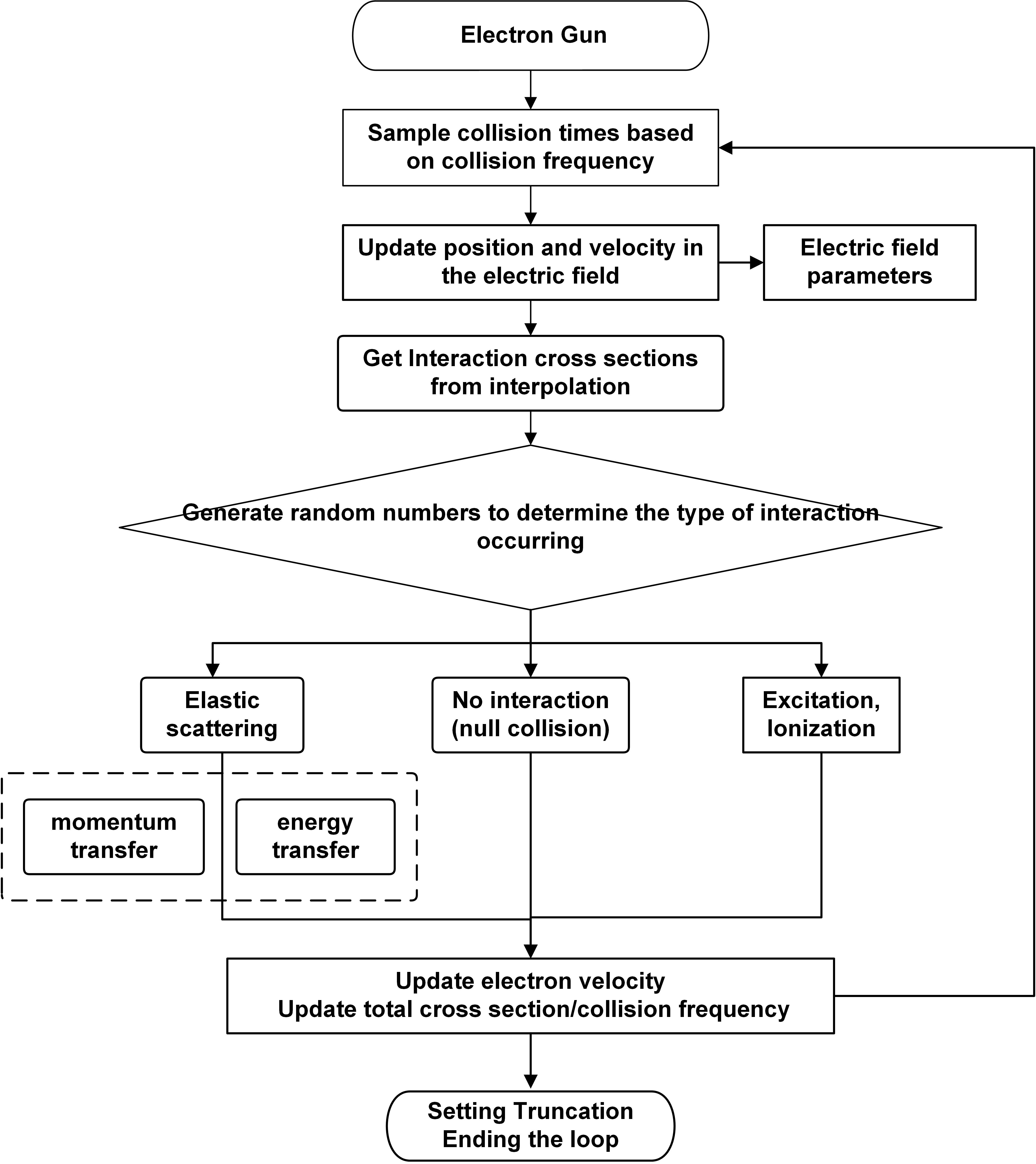}
 \DeclareGraphicsExtensions.
\caption{The framework of the Monte Carlo simulation for electron transport}
\label{fig1}
\end{figure}

The framework of Monte Carlo simulation for electron transport in an electric field is primarily based on electron-atom collisions, encompassing elastic scattering, ionization and excitation processes.
Two considerations must be made to model the collisions in the presence of an electric field. The first is that electrons accelerate in an electric field, which changes the velocity between collisions compared to having no field. The second is that the cross section for elastic scattering depends on the velocity of the interacting electrons. Relying solely on the collision frequency based on a specific initial velocity may cause bias. Two techniques, the “integral technique” and the “null collision technique”, can be used to mitigate the influence of the collision frequency selection due to the existence of an electric field.
\begin{equation}
\label{eq1}
P(\tau) = \exp\left(-\int_{0}^{\tau} \nu\left(| v + a t  |\right) dt \right)
\end{equation}
here $a$ is the electron acceleration in the electric field. $\nu$ is the collision frequency. $v$ is 
 the velocity. $P(\tau)$ is the probability of an average collision time $\tau$.

\indent  In the "integral technique", the particle's equation of motion is integrated. This technique is more difficult in terms of the mathematical treatment. In this work, we choose the "null collision technique", which involves increasing the electron collision frequency while not changing the actual interaction probability.
The essence of the "null collision technique" is to realize numerical integration of the electron velocity in the electric field between collision points. Increasing the collision frequency ($n×\nu$, $\nu$ is the original frequency and $n>1$) is equivalent to inserting additional collision points in the mean free path. At the same time, the total interaction probability of {100\%} at the end of the mean free path becomes {1/n}. In this way, more accurate simulation can be achieved.
In this study, the collision frequencies of 2x, 5x, and 10x are implemented to ascertain the appropriate collision frequency for the simulation.

The mean free path of the electron $\lambda$ and mean collision time $t$ are expressed as
\begin{equation}
\label{eq3}
\lambda = \frac{1}{n\sigma_{tot}}   ~~and~~   \nu=\frac{v}{\lambda}
\end{equation}
where $n$ is the number density of atoms. $\sigma_{tot}$ is the total cross section. $\nu$ is the collision frequency.

Using the "null collision technique", the new collision frequency $\nu'$ ,greater than $\nu$, is employed as $2\nu$, $5\nu$ and $10\nu$. Let $\nu' = 2\nu / 5\nu / 10\nu$, sampling of the collision time is:
\begin{equation}
\label{eq2}
T = \frac{-1}{\nu'} \ln(R) 
\end{equation}
$R$ is a random number between 0 and 1. $\nu'$ is the new collision frequency.  

\indent The simulation framework, depicted in Fig.\ref{fig1}, is initiated by defining the essential electron parameters given by users, including the electron position, and energy etc. Using the "null collision technique", the mean collision time is sampled, shown in Eq.\ref{eq2}. The electron is accelerated in the electric field during this time period, and the electron's position and velocity are updated. The interaction probability is then determined based on the relative velocities between the electron and the background atom. By generating a random number between 0 and 1, the type of interaction (ionization, excitation, elastic scattering, or non-interaction) is chosen by the corresponding probability interval in the random number falls.

\indent For the electron-atom interactions, excitation and ionization of background atoms occur when the electron's energy reaches the specific energy threshold.
The atom absorbs external energy and its electron transition from an orbit of lower energy to one of higher energy. Therefore, the excitation is characterized by the removal of energy at the corresponding energy level for the electron. The subsequent formation of X-rays or Auger electrons, which may result from such excitation, is not within the consideration. For the ionization,  a new electron is generated, which inherits the position of the parent electron, and is then incorporated into a new loop in Fig.1. When the electron is into a new loop, the electron's step length is computed, the interactions are sampled, and particle changes are updated.

\indent After the interaction is completed, the electron can stop running if the set criteria are met. A variety of criteria can be used to determine the termination of the loop, such as (a) the electron's movement time being more than a set time. (b) the electron's movement distance reaching a certain length, or even the electron's position in a certain direction reaching a set position. (c)  the electron's energy is lower or higher than a certain predetermined value and so on.

\subsection{Liquid phase simulation framework}

The simulation of electron transport in liquids cannot be achieved only by changing the atom density in the model.
Due to the increase in atom density from gas to liquid phase and the low energy region of interest for electrons, the electron wavelength is comparable to the atom distance. For example, for an atom density of $1e22 cm^{-3}$, the mean atom distance is approximately $N^{-1/3}\approx1 nm$, aligning with the wavelength of 1 eV electrons within the same order of magnitude. Thus coherent scattering needs to be considered. The theory of elastic scattering in liquids including coherent scattering is described in detail in the CL theory\cite{cohen1967theory} and is briefly outlined below.

\subsubsection{Coherent scattering}

 In liquids, coherent scattering has to be taken into account. An electron is considered more as waves interacting with multiple scattering centers instead of as a single particle. The fundamental attribute of liquid is the short-range order, which is distinct from the long-range periodicity characteristic of a crystalline solid. Given that the structure of crystals could be measured by X-ray Bragg reflection, it follows that the liquid structure factor S(k) can also be ascertained by the scattering experiment.\cite{march1991atomic} For example, the intensity $I$ of X-ray/neutron is measured after a ray of intensity $I_0$ has been scattered from a liquid medium across a certain angle $2\theta$.\cite{PhysRevA.7.2130}\cite{Daniell1989}

Let $k = 4\pi \sin \theta/\lambda$,  the liquid structure factor S(k) is defined as
\begin{equation}
S(k) = \frac{I}{N f^2}
\end{equation}
where $N$ is the total number of atoms in noble liquid, $\lambda$ is the ray wavelength. $f$ is the atom scattering factor, which is the Fourier transform of the electron density in the atoms.
Because the positions of the atoms are spatially correlated, the structure factor S(k) is related to the pair correlation function $g(r)$ via Eq.\ref{equ5}
\begin{equation}
\label{equ5}
S(k) = 1 + \rho \int\left[ g(r) - 1 \right] e^{(i k \cdot r)} dr
\end{equation}
where $g(r)$ is the pair distribution function. It describes the distribution of distances between atoms in a substance. Zernike and Prins\cite{zernike1927beugung} set $\rho g(r)4\pi r^2 dr=N$, and $N$ equals to the total number of atoms in a spherical shell of radius $r$ and thickness $dr$ centred on a chosen atom at the origin of coordinates, $\rho$ is the average number density of $N$ atoms in volume $V$. 

\indent In the CL theory, the electron distribution function is expanded through Legendre polynomials and brought into the electron Boltzmann equation. Subsequently, for collision terms, considering the small mass ratio between electrons and atoms and the properties of $S(k)$, the mean free path of electron elastic collisions in the liquid phase is divided into two types: energy transfer and momentum transfer.
The energy transfer mean free path $\Lambda_{0}$ is as follows:

\begin{align}
\Lambda_{0}(\epsilon)^{-1} &= N 2\pi\int_0^\pi \Sigma(\theta, \epsilon) (1 - P_i(\cos\theta)) \sin\theta d\theta \\&= N \Sigma(\epsilon) 
\end{align}
The momentum transfer mean free path $\Lambda_{1}$ is as follows:
\begin{align}
\Lambda_{1}(\epsilon)^{-1} &= N 2\pi\int_0^\pi \Sigma(\theta, \epsilon) (1 - P_i(\cos\theta)) S(K) \sin\theta d\theta \\&= N \Sigma(\epsilon) S(K)
\end{align}
where $\epsilon$ is the electron energy. $P_i(\cos\theta)$ is the Legendre polynomial. $\Sigma(\theta, \epsilon)$ is the differential scattering cross section of the electron. The equation integrates over the angle of the differential cross section $\Sigma(\theta, \epsilon)$ to obtain the scattering cross section $\Sigma(\epsilon)$. $ S(K) $ is the structure factor discussed earlier.

\subsubsection{Elastic scattering simulation model}

\begin{figure}[!t]
\centering
\includegraphics[width=\textwidth]{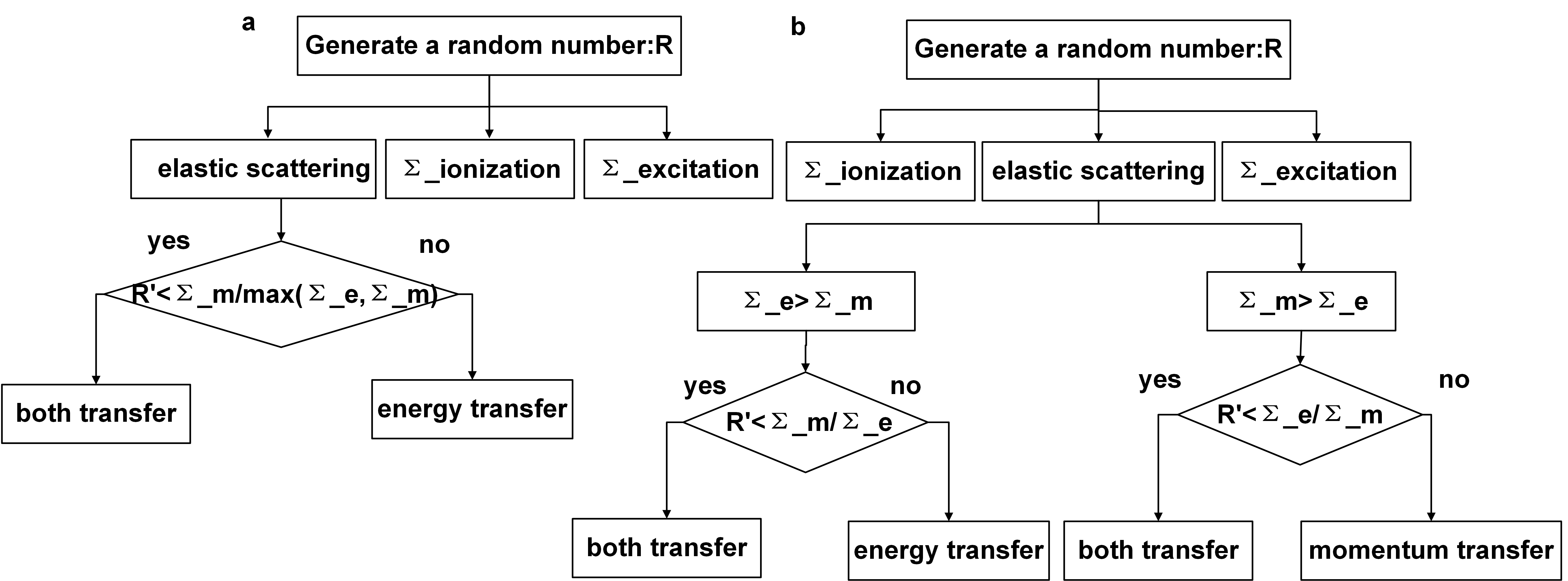}
 \DeclareGraphicsExtensions.
\caption{ Monte Carlo elastic scattering sampling model in liquid, the left is the WT model and the right is SSMC model }
\label{figliquidmodal}
\end{figure}

\indent In the simulation, according to the different frequencies of energy and momentum transfer, the elastic scattering includes three parts:

\noindent(a) Energy transfer: change the electron's energy without changing the direction. 

\noindent(b) Momentum transfer: change the direction of the electron without changing the energy. 

\noindent(c) Both energy and momentum transfer (single-particle collision): Update the electron's energy and direction in accordance with the mass ratio following the elastic scattering principle between an electron and an atom.

 \indent There are WT and SSMC models for the electron elastic scattering simulation in liquids, as shown in Fig.\ref{figliquidmodal}. 
 The energy transfer or both transfer in WT model is determined by the ratio of $\Sigma_e $ and $\Sigma_m $. $\Sigma_e$ is the elastic energy-transfer cross section and $\Sigma_m$ the momentum-transfer cross section.
 Since both transfer has energy and momentum transfer, the frequency of both transfer increases as $\Sigma_m /\Sigma_e $ goes up.
 For the SSMC model, the difference between the two is in the region where $\Sigma_m$ is larger than $\Sigma_e $. The SSMC model treats the energy/momentum transfer mode separately based on the $\Sigma_e $ and $\Sigma_m $ size relationship, emphasizing the importance of momentum transfer in the region of $\Sigma_m > \Sigma_e $. For noble liquids, since the region of $\Sigma_m > \Sigma_e $ is small, both models can be used for electron transport in noble liquids.

\section{Result and Discussion}
\label{section 3}
\subsection{Swarm parameters in simulation}
\indent The result from the electron transport Monte Carlo simulation can be verified by comparing the electron swarm parameters with the experimental or theoretical results including the electron drift velocity and diffusion coefficient. 

\begin{figure}[!ht]
\centering
\includegraphics[width=0.9\textwidth]{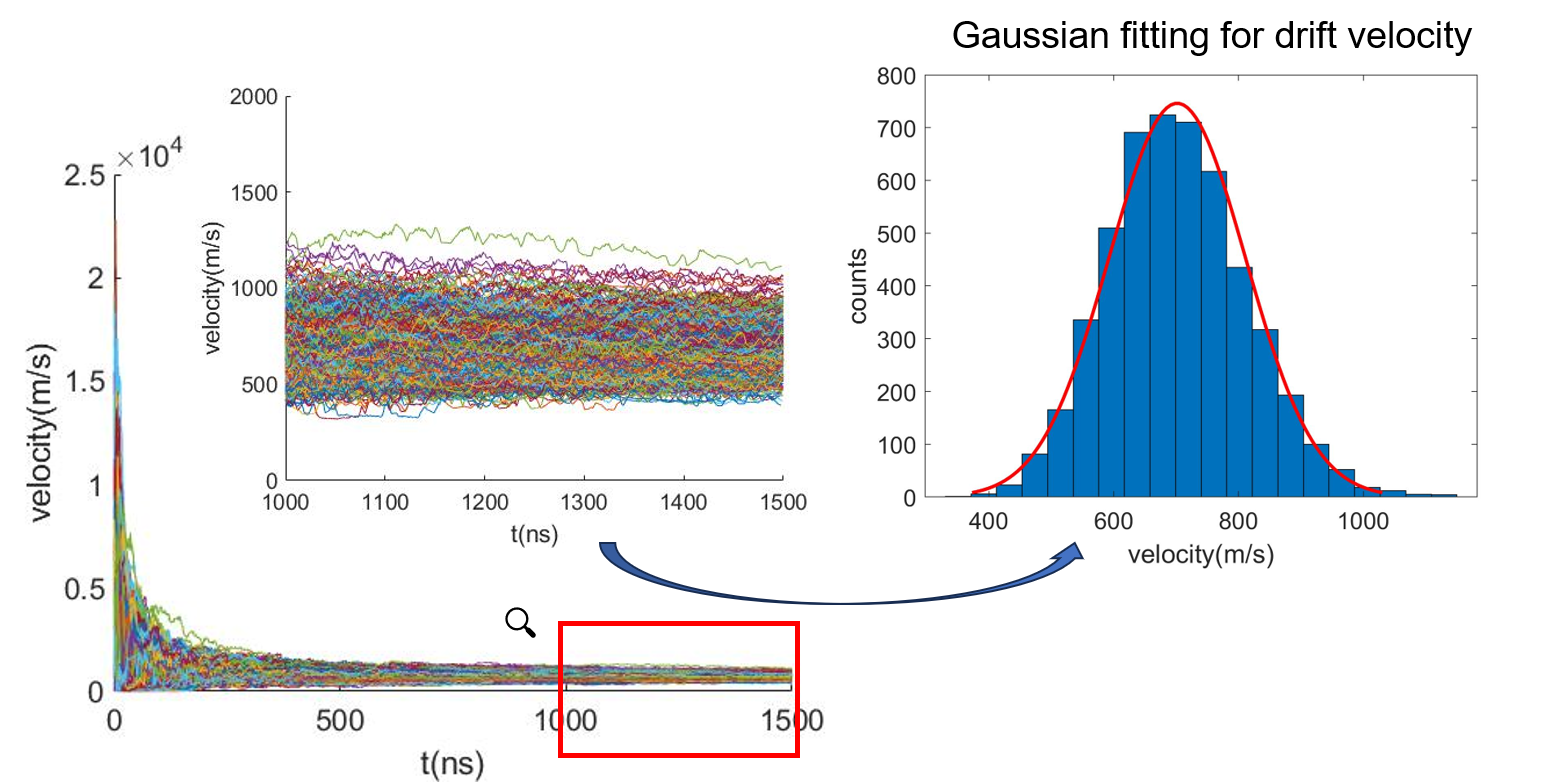}
 \DeclareGraphicsExtensions.
\caption{ Electron drift velocity over time in an electric field of 100 V/cm at 295 K, with demonstrated local zoom details and Gaussian fit to drift velocity.}
\label{速度}
\end{figure}

The electron drift velocity is obtained as the ratio of the distance in the electric field direction to the total drift time. Taking 5000 electrons with the initial energy of 1 eV at 295 K as an example, Fig.\ref{速度} depicts the drift velocity of the electrons over time. Since the electron's initial velocity is higher than the thermal equilibrium velocity, the electron velocity experiences a rapid decrease on the order of nanoseconds before it reaches a stable drift velocity. 
The simulated drift velocity is determined by a Gaussian fit to the drift velocity data after a number of electrons have stabilized. The mean value obtained from the fit is the final simulated drift velocity.

\begin{figure*}[!ht]
\centering
\subfloat[longitudinal and transverse diffusion]{\includegraphics[width=0.49\textwidth]{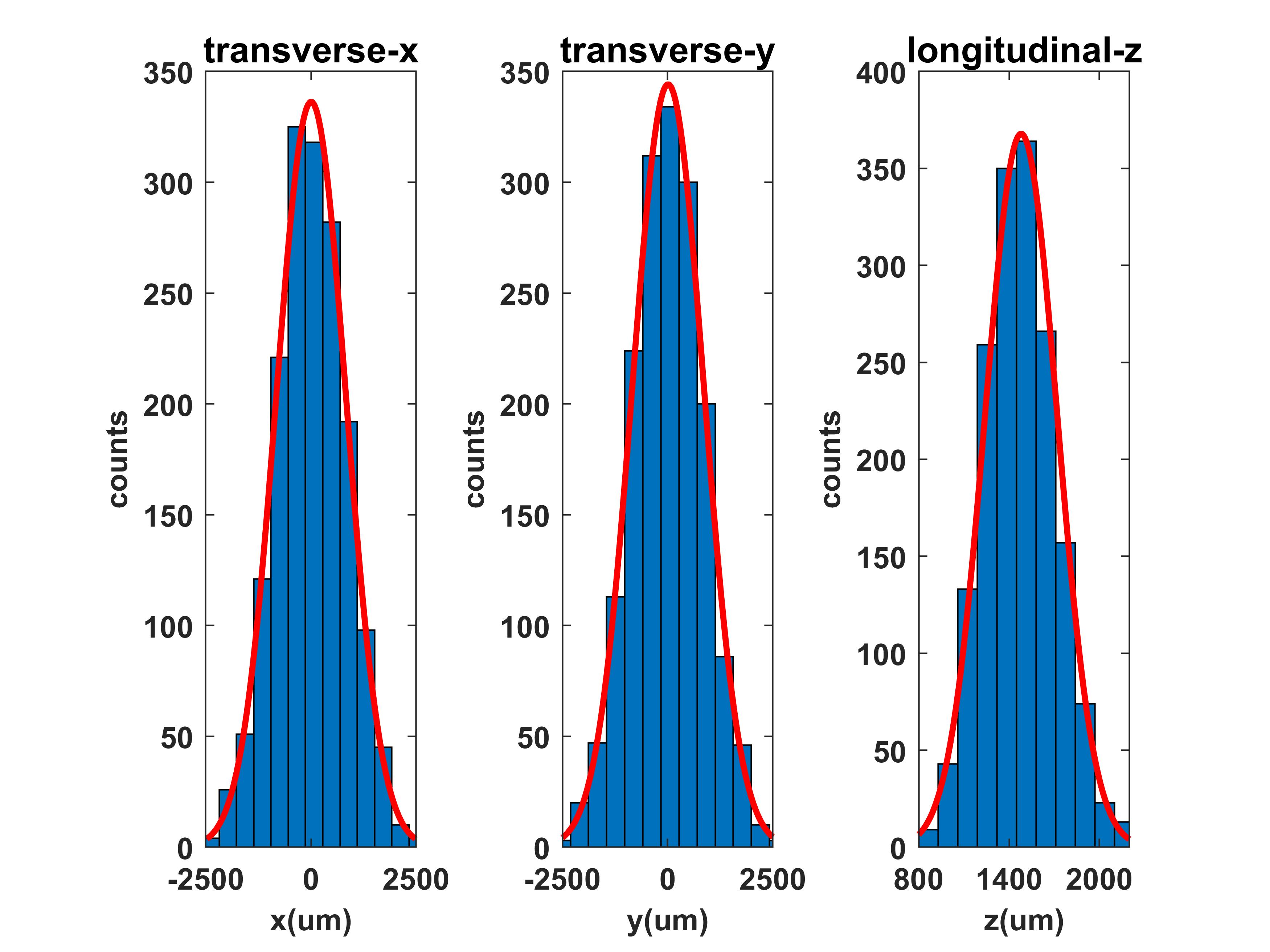}}
\hfil
\subfloat[Gaussian fit variance over time]{\includegraphics[width=0.49\textwidth]{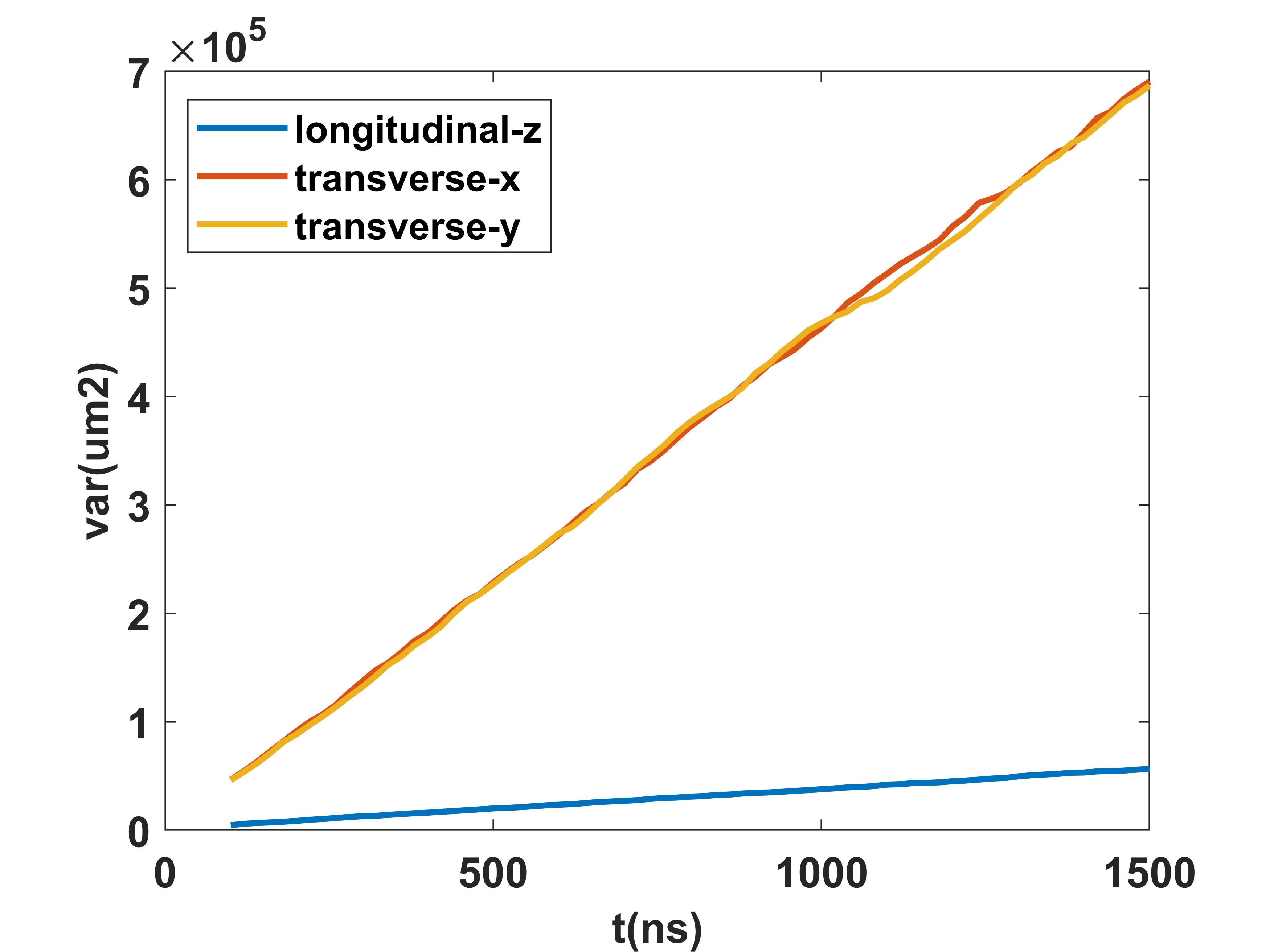}}
 \caption{the transverse and longitudinal diffusion of electrons in gas argon. (a) The position distribution of the electrons is shown at 1500 us. Z is the direction of the electric field. X and y are the transverse direction. (b) is a linear fit of the variance over time, with the diffusion coefficient being this slope divided by 2. }
\label{diffusion-figure}
\end{figure*}

The diffusion coefficient of electrons is also one of the key parameters resulting from collisions with the medium's atoms. 
The diffusion coefficient $D$ quantifies extent of the electron cloud's diffusion in the substance. The electron frequency distribution concerning distance $r$ and drift time $t$ is shown in Eq.\ref{electron frequency function}\cite{Schmidt1997LiquidSE}. 
By analyzing the three-dimensional positions of electrons at time $t_i$, Gaussian fits are applied to distances along ($z$) and perpendicular ($x,y$) to the electric field direction to determine the standard deviation ($\sigma_{xi}$, $\sigma_{yi}$, $\sigma_{zi}$), as shown in Fig.\ref{diffusion-figure}. Subsequently, according to Eq.\ref{diffusion}, $\sigma_{i}^2$ is linearly fitted at time $t_i$ and the fitting slope divided by $2$ is the final diffusion coefficient ($D_L$, $D_T=\frac{D_{Tx}+D_{Ty}}{2}$).

\begin{equation}
f(r, t) = \frac{N_0}{4 \pi Dt} \exp{\frac{-r^2}{4Dt}}
\label{electron frequency function}
\end{equation}
where $N_0$ is the total number of electrons. As the Gaussian distribution:
\begin{equation}
\sigma=\sqrt{2Dt}
\end{equation}

\begin{equation}
D=\frac{\sigma^2}{2t} 
\label{diffusion}
\end{equation}

\subsection{Gas phase}

\begin{figure}[!ht]
\centering
\includegraphics[width=\textwidth]{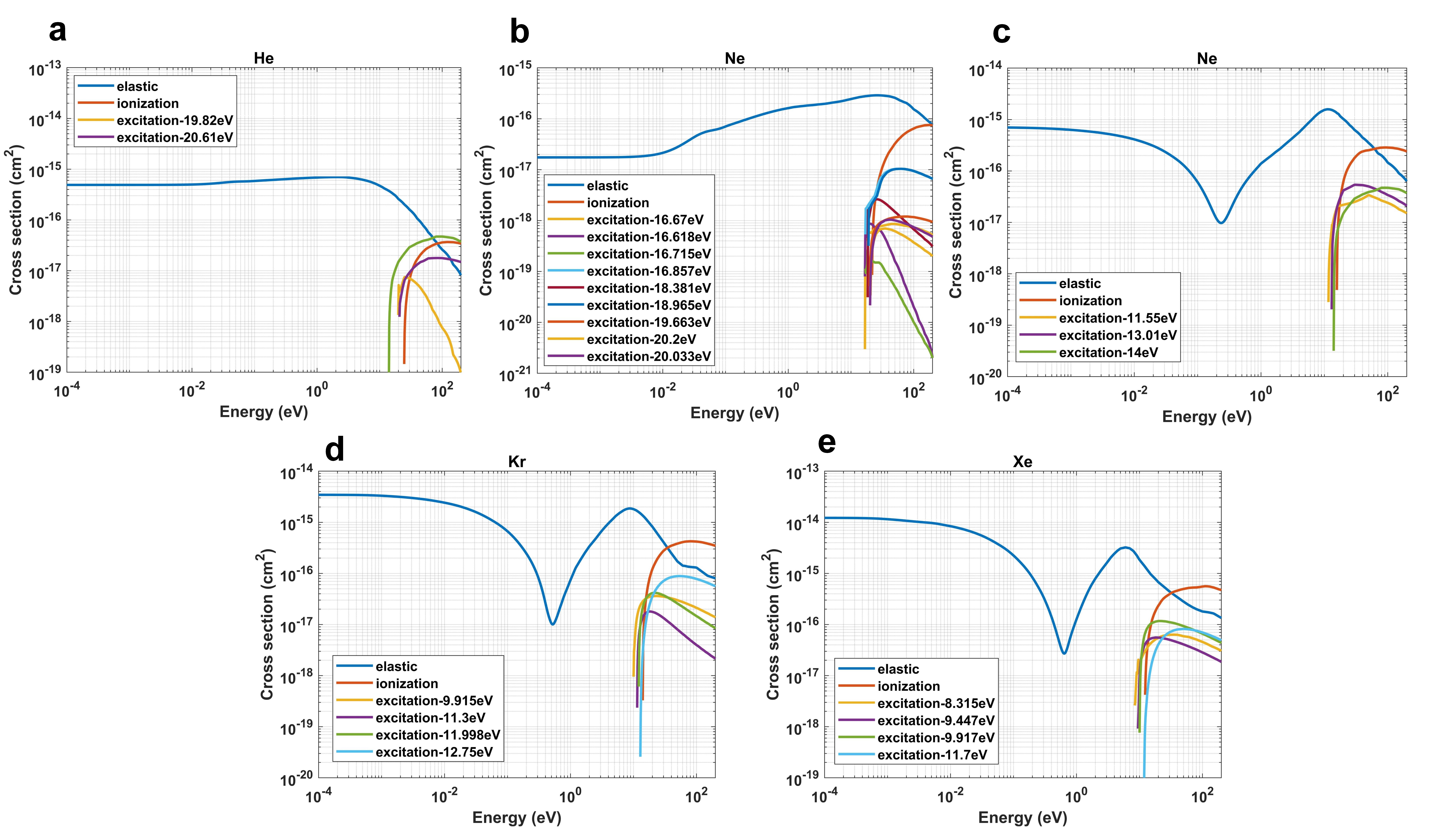}
 \DeclareGraphicsExtensions.
\caption{ Elastic scattering, excitation and ionization cross sections for noble gas(from Biagi LXCat\cite{https://doi.org/10.1002/ppap.201600098}\cite{carbone2021data}\cite{PANCHESHNYI2012148}). a:Helium, b:Neon, c:Argon, d:Krypton, e:Xenon.}
\label{fig2}
\end{figure}

\begin{figure}[h]
\centering
\subfloat[The electron drift velocity in gas He]{\includegraphics[width=0.45\textwidth]{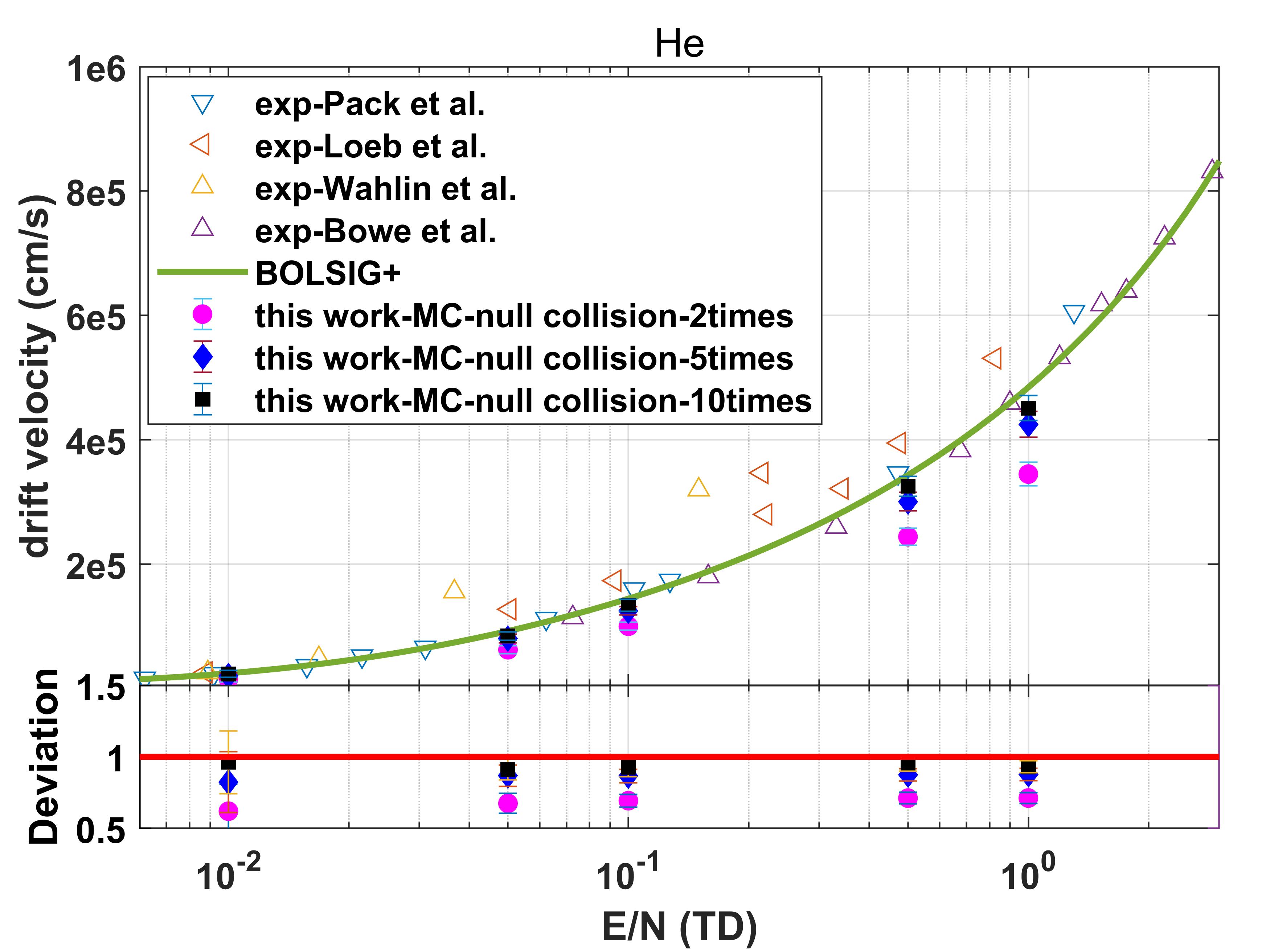}}
\label{fig_sub1.1}
  \hfil
\subfloat[The electron diffusion in gas He]{\includegraphics[width=0.45\textwidth]{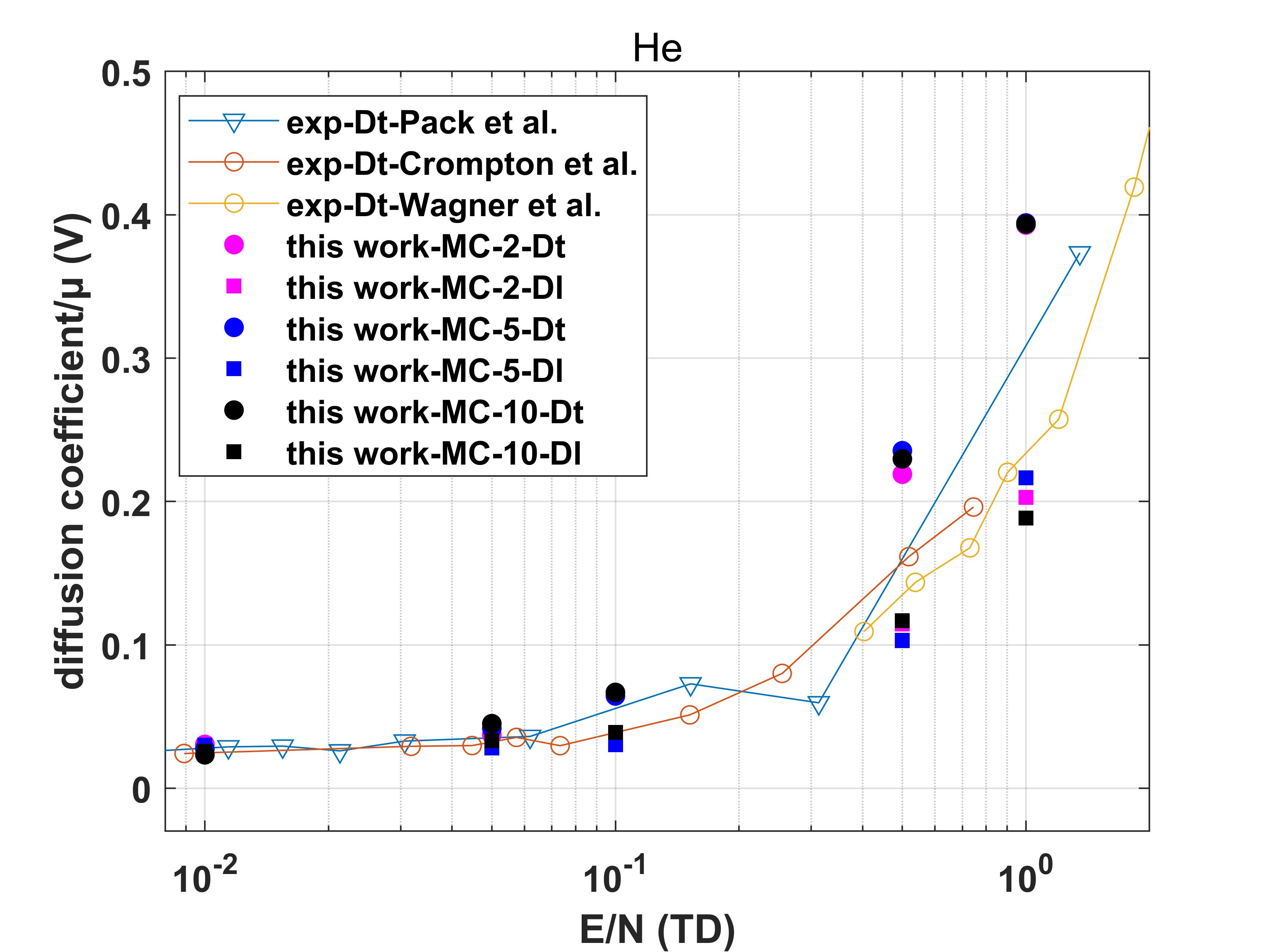}}

\label{fig_sub1.2}
 \caption{Electron swarm parameters in gas helium. The hollow symbols represent experimental data and are labeled as "exp". The solid symbols indicate simulation results, and the line represents theoretical results. (a) The electron drift velocity varies with the reduced electric field $E/N$. The experimental drift velocity data from Pack\cite{pack1961drift}, Loeb\cite{loeb1924mobilities}, Wahlin\cite{wahlin1926motion}, Bowe\cite{bowe1960transport} and the data from BOLSIG+(a Boltzmann Equation Solver)\cite{Hagelaar2005SolvingTB} are compared. The deviations between the simulation results and BOLSIG+ are shown at the bottom of the figure. (b) The ratio of the electron diffusion coefficient to mobility varies with the reduced electric field. The experimental data from Pack\cite{pack1992longitudinal}, Crompton\cite{crompton1967momentum}, Wagner\cite{wagner1967time} are compared.}
   \label{fig:combined1}
     
    \end{figure}
  
    
    \begin{figure}[!h]
\centering
\subfloat[the electron drift velocity in gas Ne]{\includegraphics[width=0.45\textwidth]{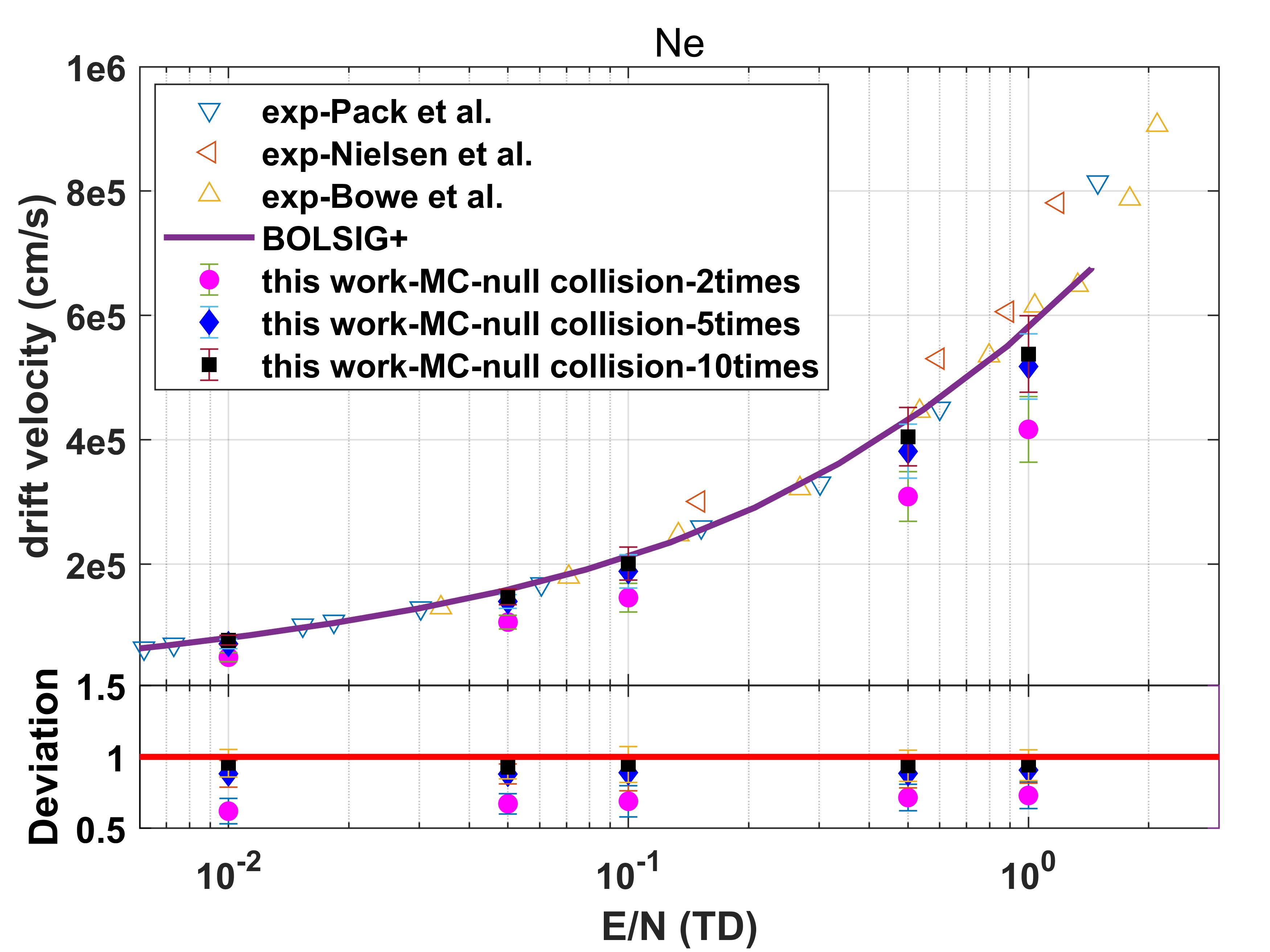}}
\label{fig_sub2.1}
  \hfil
\subfloat[the electron diffusion in gas Ne]{\includegraphics[width=0.45\textwidth]{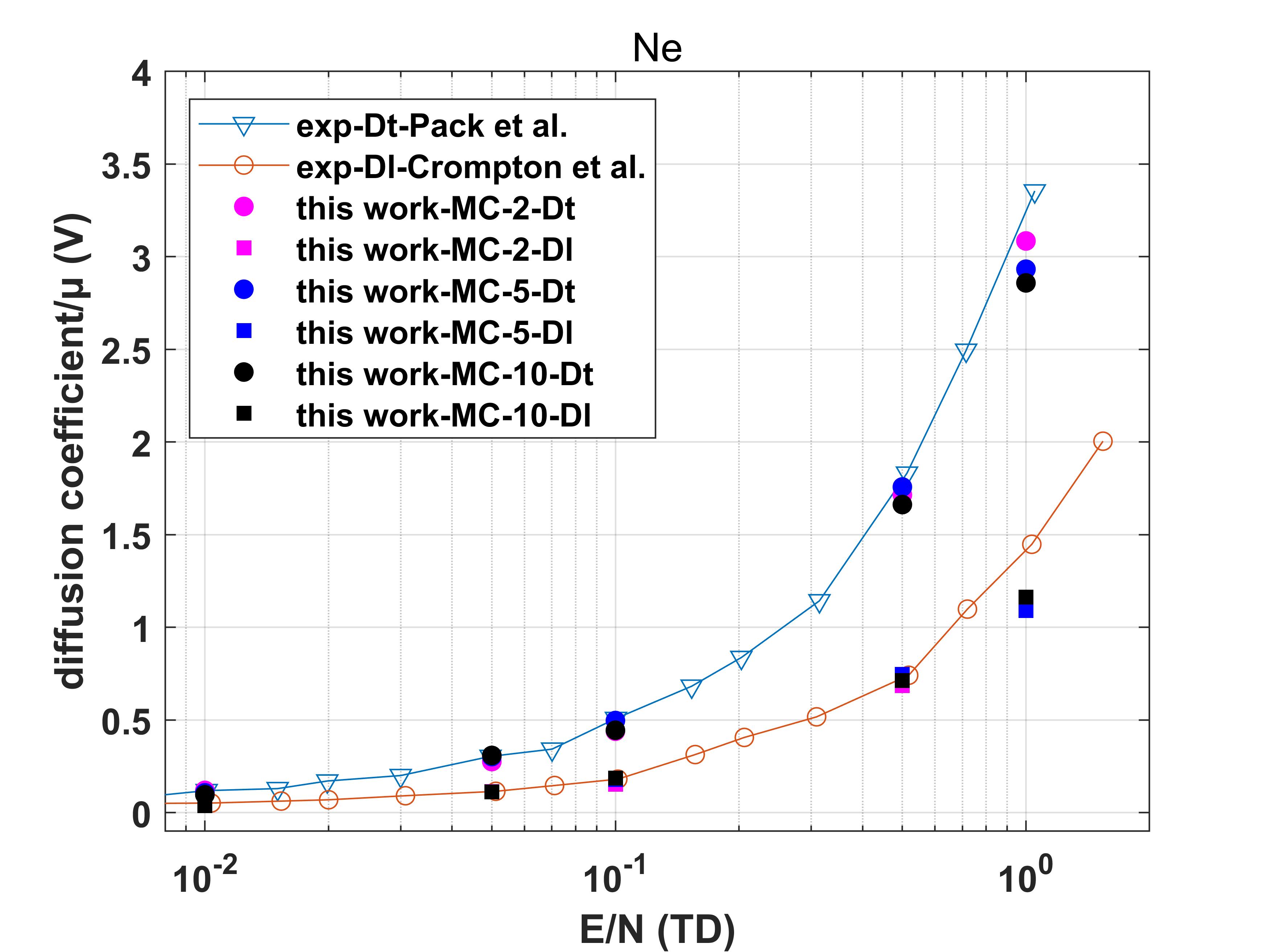}}
\label{fig_sub2.2}
 \caption{Electron swarm parameters in gas neon. The hollow symbols represent experimental data and are labeled as "exp". The solid symbols indicate simulation results, and the line represents theoretical results. 
 (a) The electron drift velocity varies with the reduced electric field. The experimental data from Pack\cite{pack1961drift}, Nielsen\cite{nielsen1936absolute}, Bowe\cite{bowe1960transport} and calculated data from BOLSIG+(a Boltzmann Equation Solver)\cite{Hagelaar2005SolvingTB} are compared. The deviation between the simulation results and BOLSIG+ is shown at the bottom of the figure. (b) The ratio of the electron diffusion coefficient to mobility varies with the reduced electric field. The experimental data from  Mayorov\cite{mayorov2021effect} are compared.}
   \label{fig:combined2}
     
 \end{figure}

 \begin{figure}[!h]
\centering
\subfloat[the electron drift velocity in gas Ar]{\includegraphics[width=0.45\textwidth]{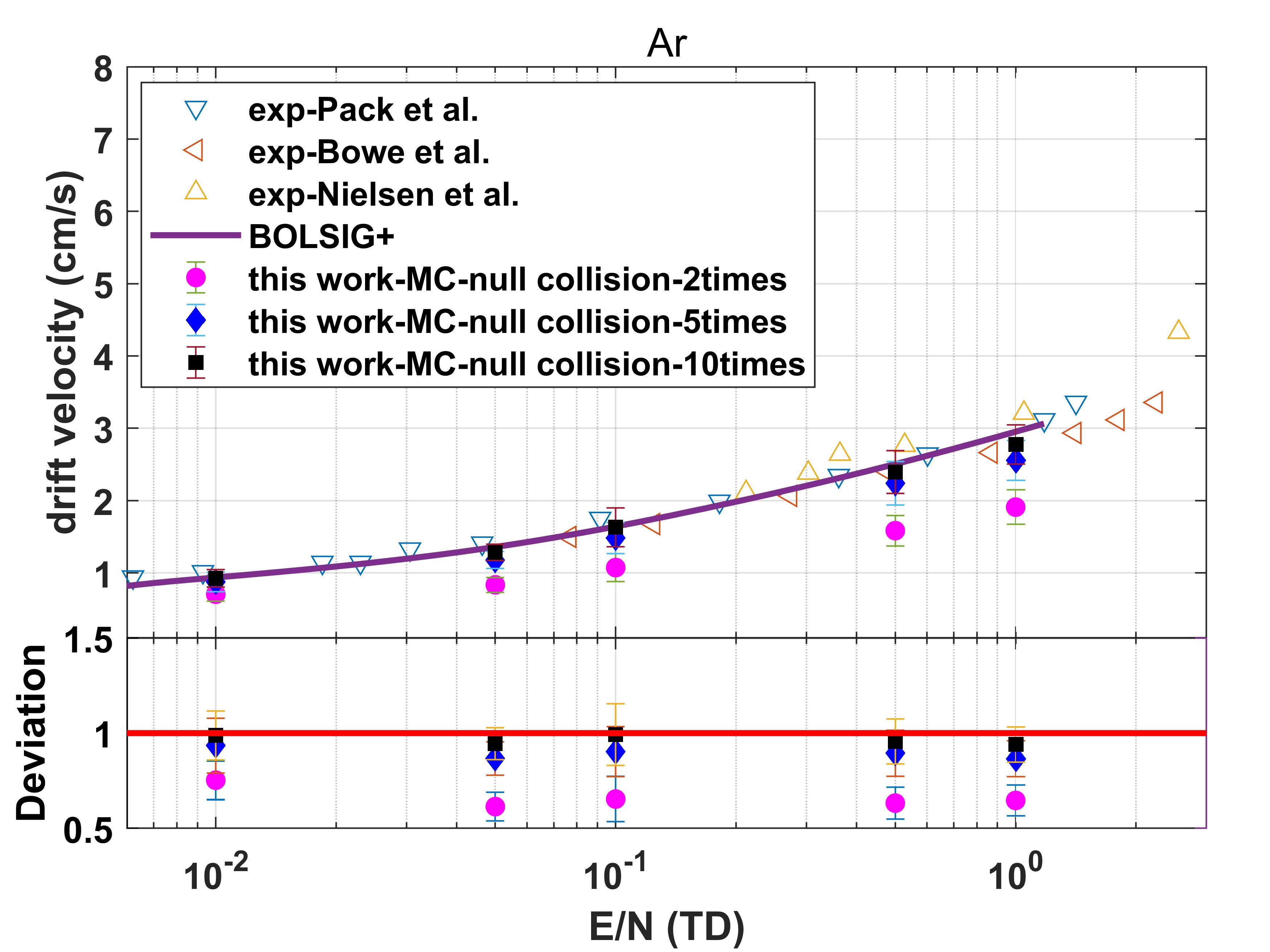}}
\label{fig_sub3.1}
  \hfil
\subfloat[the electron diffusion in gas Ar]{\includegraphics[width=0.45\textwidth]{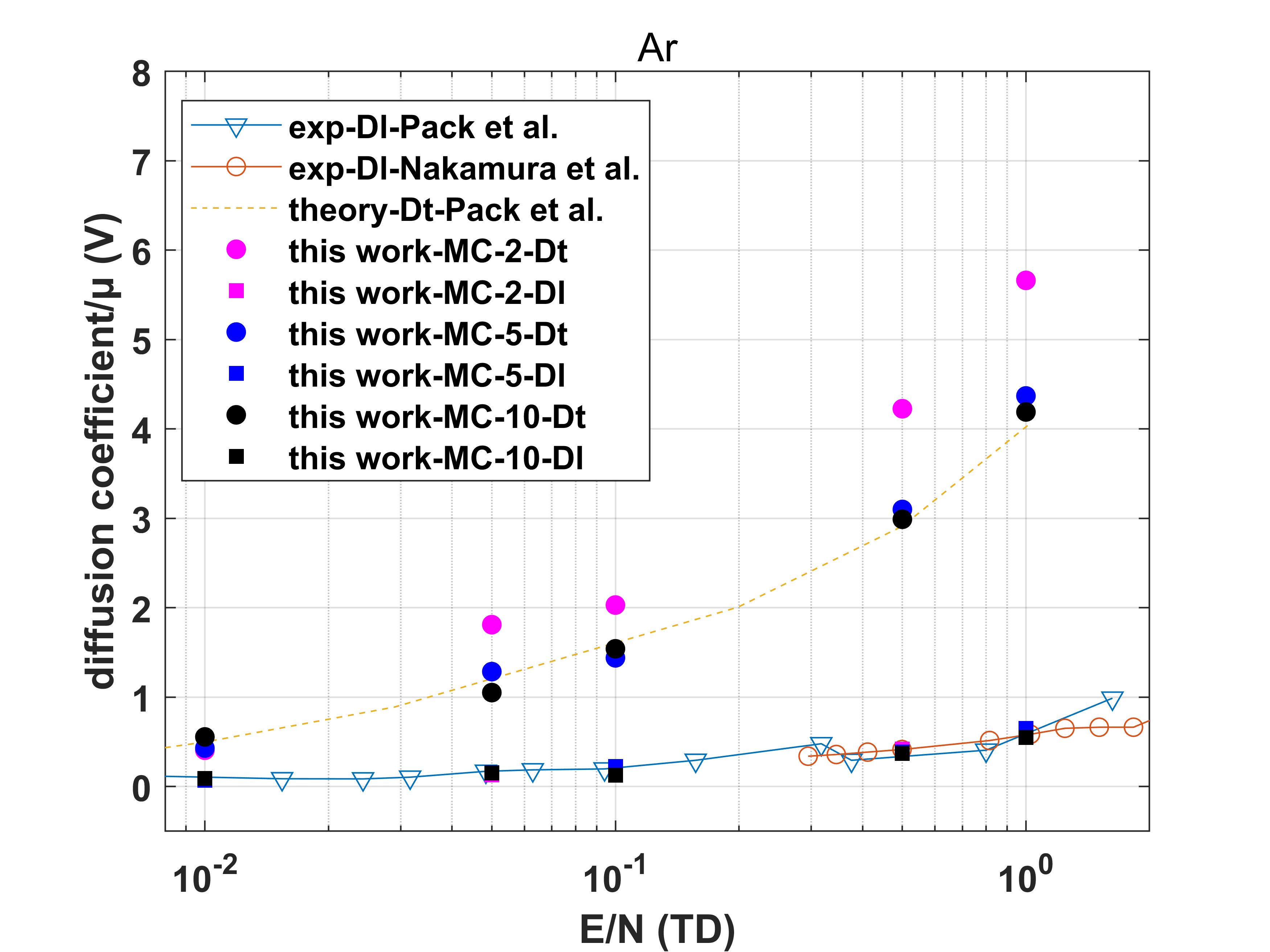}}
\label{fig_sub3.2}
 \caption{Electron swarm parameters in gas Argon. The hollow symbols represent experimental data and are labeled as "exp". The solid symbols indicate simulation results, and the line represents theoretical results. (a) The electron drift velocity varies with the reduced electric field. The experimental data from Pack\cite{pack1961drift}, Nielsen\cite{nielsen1936absolute}, Bowe\cite{bowe1960transport} and data from BOLSIG+\cite{Hagelaar2005SolvingTB} are compared. (b) The ratio of the electron diffusion coefficient to mobility varies with the reduced electric field. The experimental and theoretical data from Pack\cite{pack1992longitudinal}, Nakamura\cite{nakamura1988electron} are compared.}
   \label{fig:combined3}
 \end{figure}

 \begin{figure}[!h]
\centering
\subfloat[the electron drift velocity in gas Kr]{\includegraphics[width=0.45\textwidth]{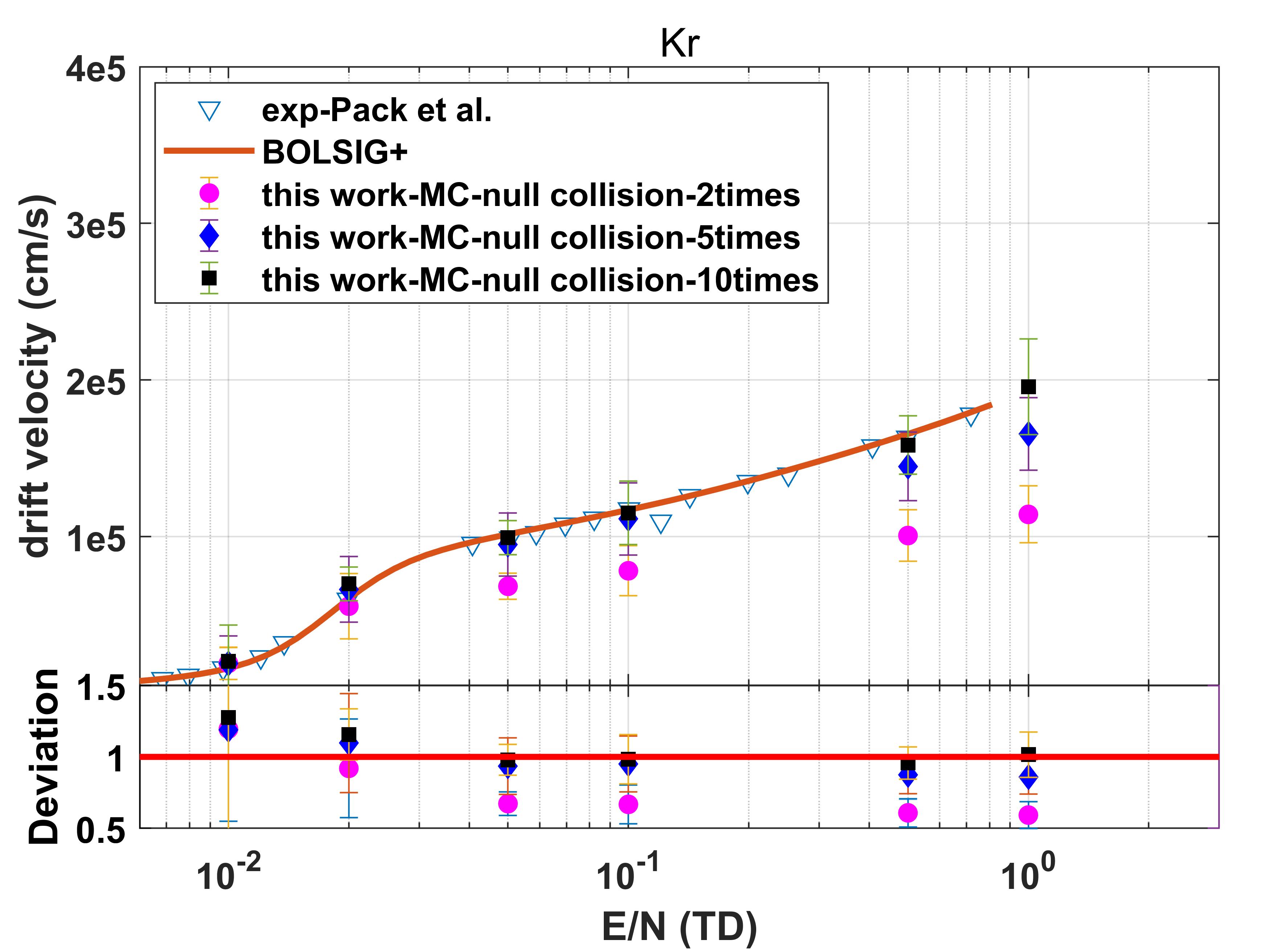}}
\label{fig_sub4.1}
  \hfil
\subfloat[the electron diffusion in gas Kr]{\includegraphics[width=0.45\textwidth]{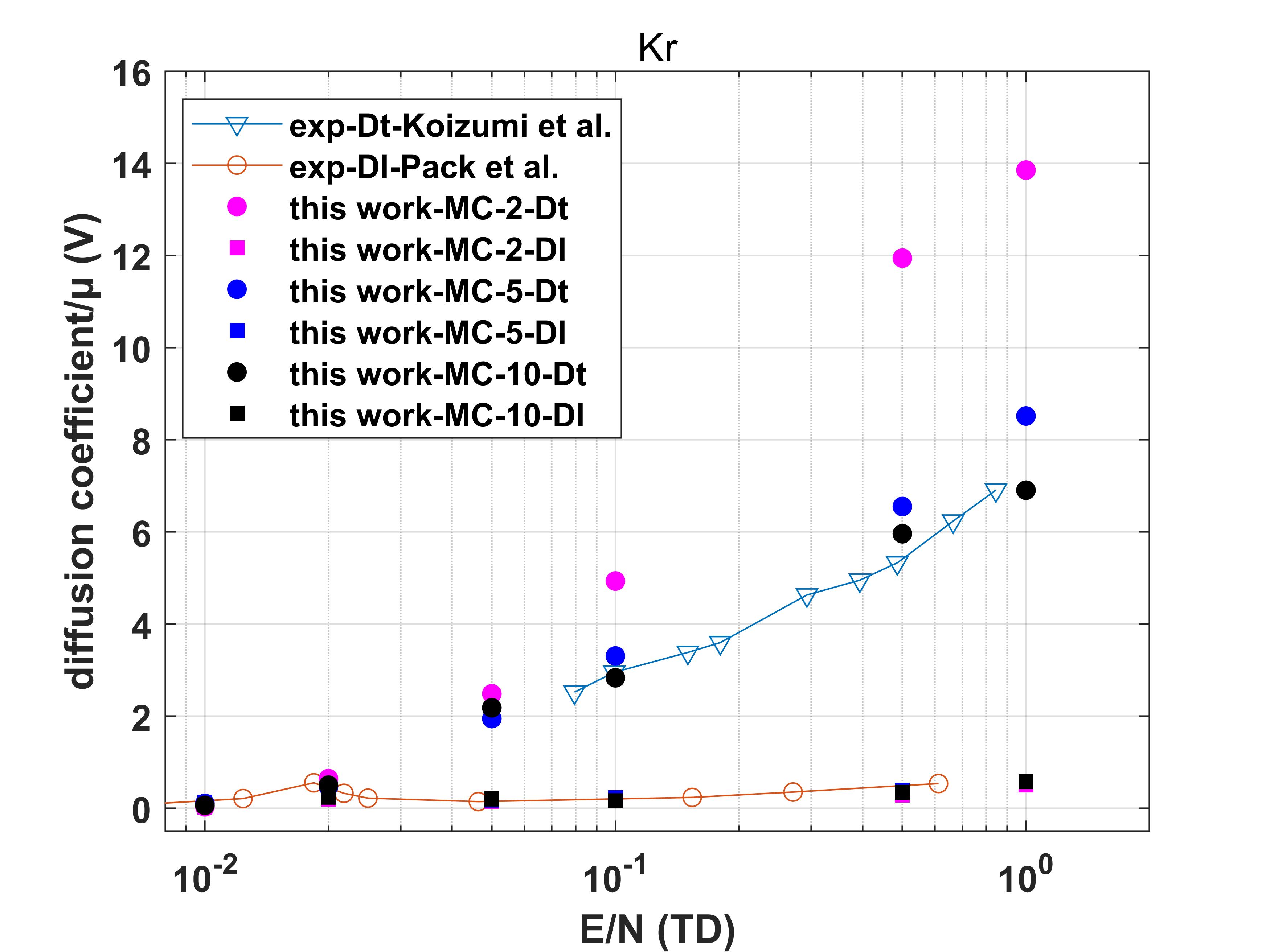}}
\label{fig_sub4.2}
 \caption{Electron swarm parameters in gas Krypton. The hollow symbols represent experimental data and are labeled as "exp". The solid symbols indicate simulation results, and the line represents theoretical results. (a) The electron drift velocity varies with the reduced electric field. The experimental data from Pack\cite{pack1961drift}, Bowe\cite{bowe1964mobility}, Brooks\cite{brooks1982electron}, English\cite{english1953grid} and data from BOLSIG+\cite{Hagelaar2005SolvingTB} are compared. (b) The ratio of the electron diffusion coefficient to mobility varies with the reduced electric field. The experimental and theoretical data from  Koizumi\cite{koizumi1986momentum} and Pack\cite{pack1992longitudinal} are compared.}
   \label{fig:combined4} 
\end{figure}

\begin{figure*}[!h]
\centering
\subfloat[the electron drift velocity in gas Xe]{\includegraphics[width=0.45\textwidth]{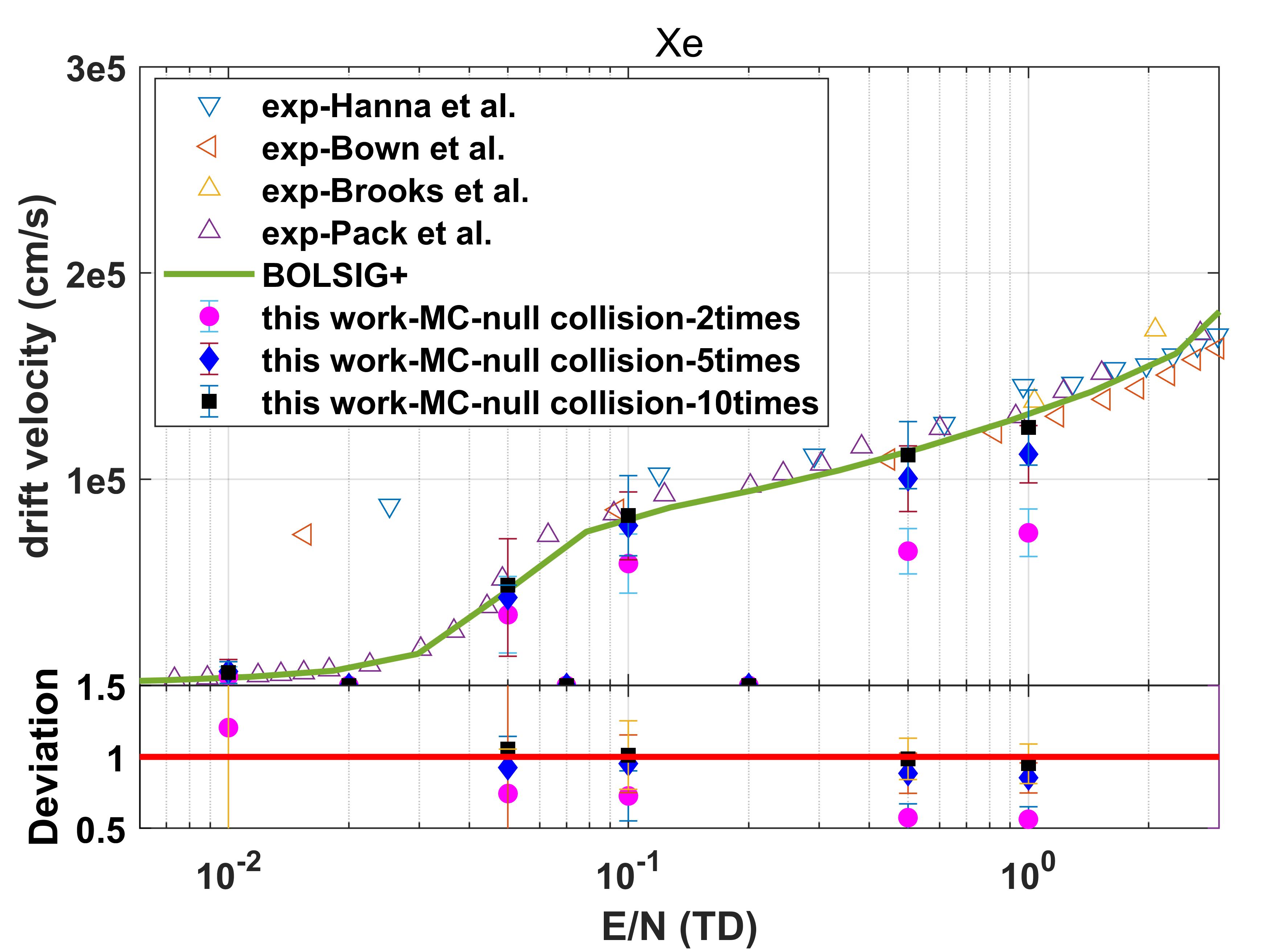}}
\label{fig_sub5.1}
  \hfil
\subfloat[the electron diffusion in gas Xe]{\includegraphics[width=0.45\textwidth]{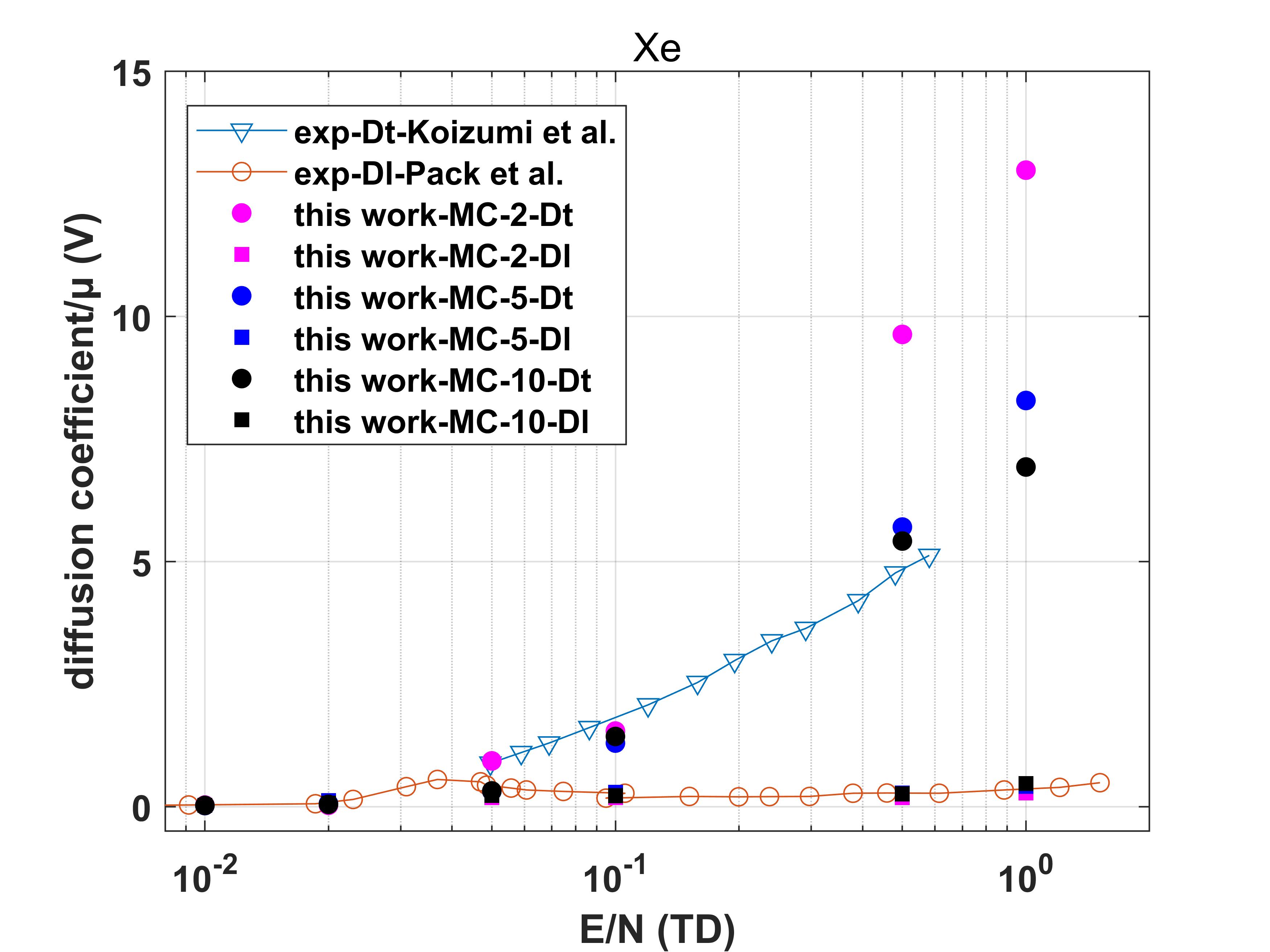}}
\label{fig_sub5.2}

 \caption{Electron swarm parameters in gas Xenon. The hollow symbols represent experimental data and are labeled as "exp". The solid symbols indicate simulation results, and the line represents theoretical results. (a) The electron drift velocity varies with the reduced electric field. The experimental data from English \cite{english1953grid}, Bown\cite{bowe1964mobility}, Brooks\cite{brooks1982electron}, Pack\cite{pack1961drift} and data from BOLSIG+\cite{Hagelaar2005SolvingTB} are compared. (b) The ratio of the electron diffusion coefficient to mobility varies with the reduced electric field. The experimental and theoretical data from  Koizumi\cite{koizumi1986momentum}, Pack\cite{pack1992longitudinal} are compared.}
   \label{fig:combined5}
\end{figure*}

\indent In gas phase, the interaction cross sections of the gas helium, neon, argon, krypton, xenon are taken from LXCat\cite{lxcat} including elastic scattering, excitation and ionization cross sections, shown in Fig.\ref{fig2}.
The atom density of all gases is $1e20 cm^{-3}$ with the temperature at 279K. The simulated drift velocity, transverse, and longitudinal diffusion coefficients for noble gases are shown in Fig.\ref{fig:combined1}-\ref{fig:combined4}. 

\indent The results of the simulated drift velocity demonstrate that as the electric field intensity increases, the electron drift velocity rises. For the drift velocity, the error bars depicted in the simulation correspond to the standard deviation of the drift velocities obtained from the ensemble of simulated electrons. Considering the effect of the "null collision technique", the simulation results are closer to the experimental results with the increase of the set collision frequency of 2x, 5x, 10x. Notably, the magnitude of the change from 5x to 10x is significantly smaller than the change from 2x to 5x. Since the higher the collision frequency setting, the longer the simulation runtime, opting for a collision frequency of 10x strikes a balance between simulation efficiency and accuracy. 

\indent In terms of diffusion, the diffusion coefficient is aligned with the experimental data. The effect of the electric field on the longitudinal diffusion coefficient is related to the trend of momentum transfer cross section with the change of energy. When the collision frequency increases with the electron's energy, the longitudinal diffusion coefficient {$D_l$} is smaller than the transverse diffusion coefficient {$D_t$}. On the contrary, when the collision frequency decreases with the electron's energy, the longitudinal diffusion coefficient {$D_l$} is greater than the transverse diffusion coefficient {$D_t$}.\cite{H R Skullerud_1969} In the range of electric field from 10V/cm to 2000V/cm, the thermal equilibrium energy of electrons is on the order of meV. As shown in Fig.\ref{fig2}, for helium (He) and neon (Ne), the momentum transfer cross-sections increase gradually with energy, so {$D_l$} is less than {$D_t$}. In contrast, for argon (Ar), krypton (Kr), and xenon (Xe), the momentum transfer cross-sections exhibit a decrease followed by an increase due to the Ramsauer effect. For Ar, the transition point is below 0.01td (10V/cm), so {$D_l$} is less than {$D_t$} in the electric field from 10V/cm to 2000V/cm. For Kr and Xe, the transition points occur at approximately 0.2td and 0.4td for {$D_l$}, respectively. This behavior is consistent with the observed reduction in longitudinal diffusion due to the electric field.

\indent Overall, the simulation results are in good agreement with the experimental and theoretical publications, affirming the validity of the framework for noble gases.

\subsection{Liquid Phrase}

\begin{figure*}[!ht]
\centering
\subfloat[liquid Ar]{\includegraphics[width=0.3\textwidth]{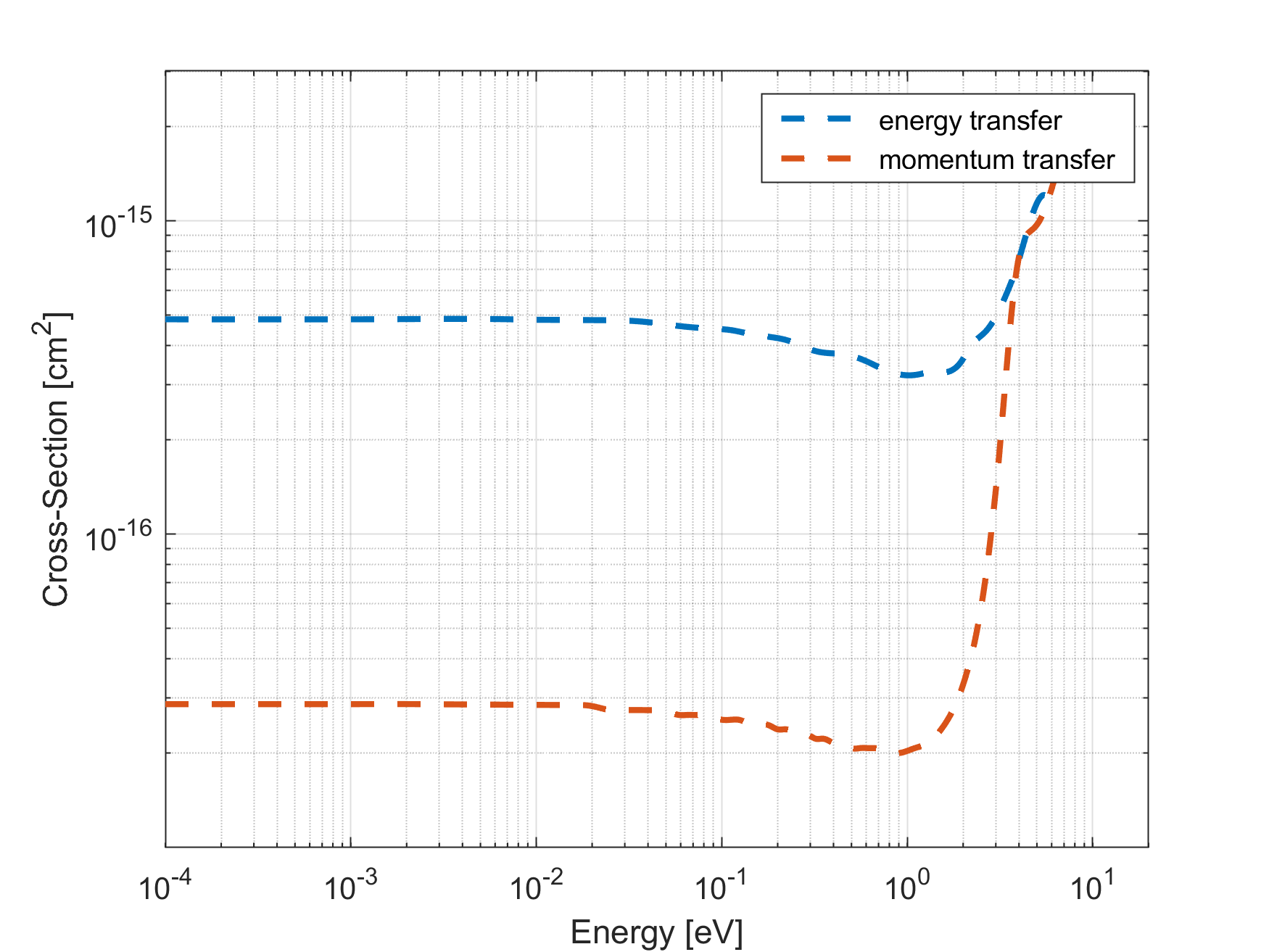}}
\label{fig_sub11.1}
  \hfil
\subfloat[liquid Kr]{\includegraphics[width=0.3\textwidth]{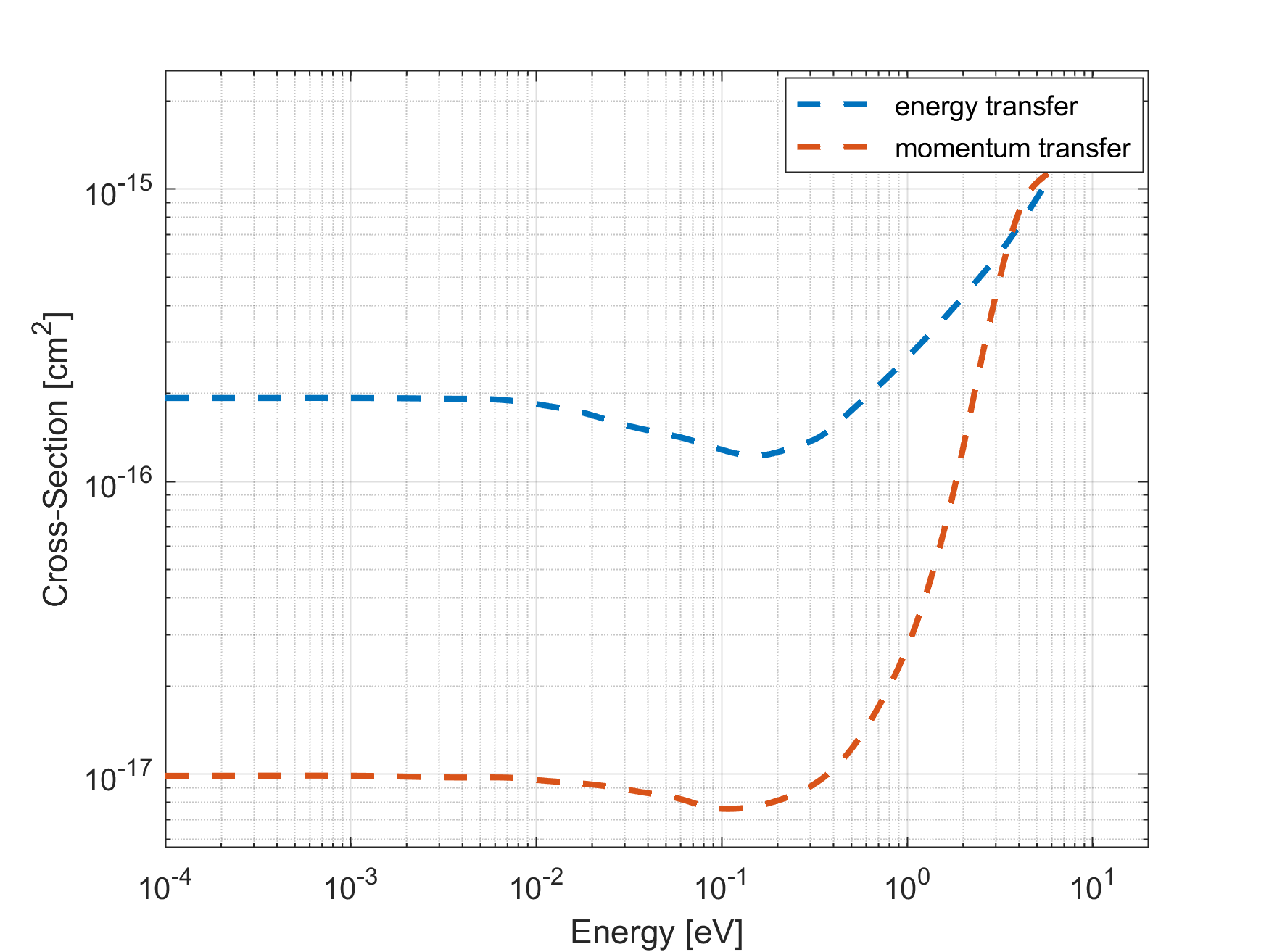}}
\label{fig_sub11.2}
  \hfil
\subfloat[liquid Xe]{\includegraphics[width=0.3\textwidth]{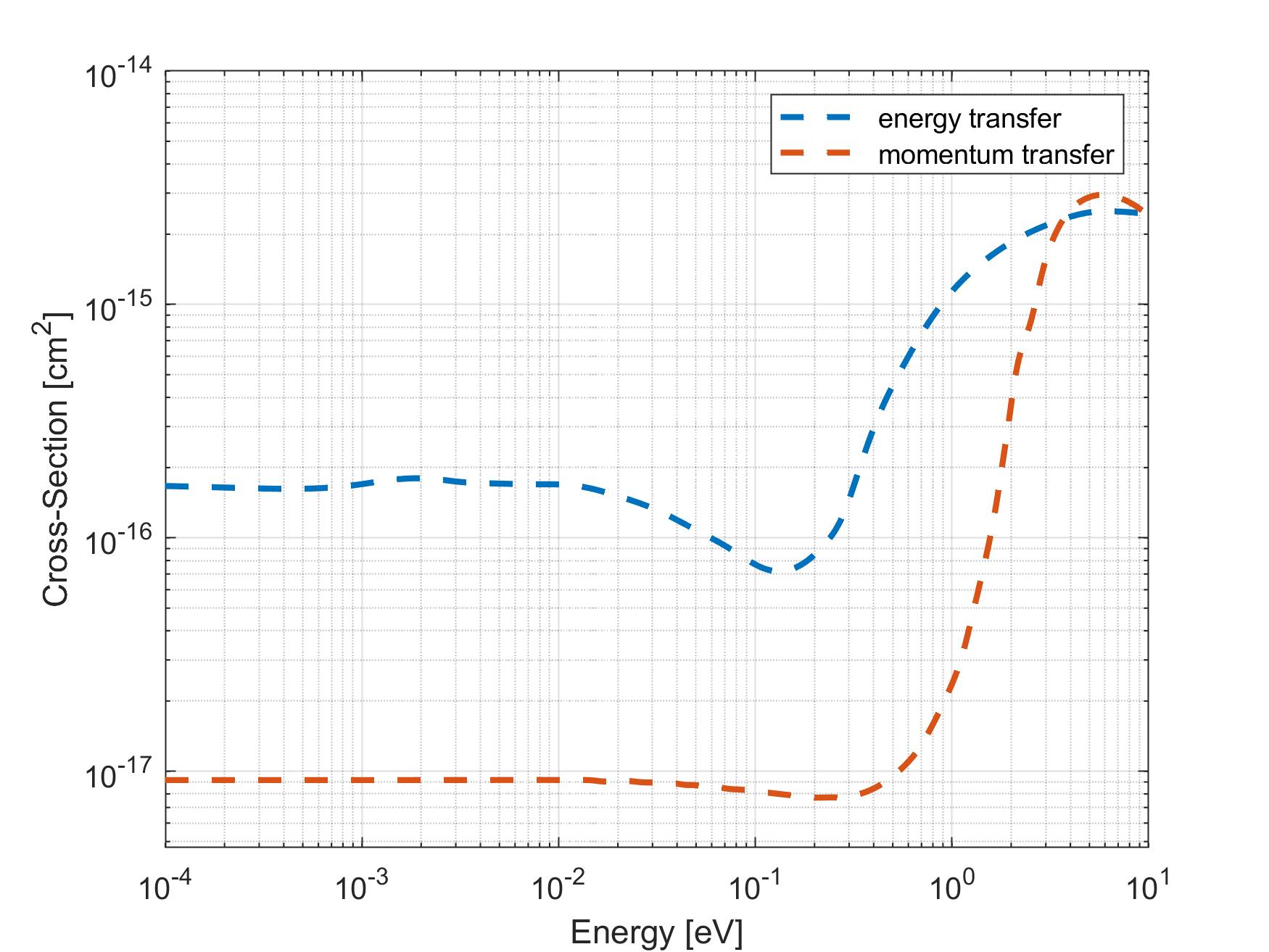}}
\label{fig_sub11.3}
 \caption{ Energy transfer and momentum transfer cross sections for noble liquids from Y.Sakai\cite{4156740}.}
   \label{cross section in liquid}
\end{figure*}

\begin{figure*}[!ht]
\centering
\subfloat[ drift velocity in LKr]{\includegraphics[width=0.45\textwidth]{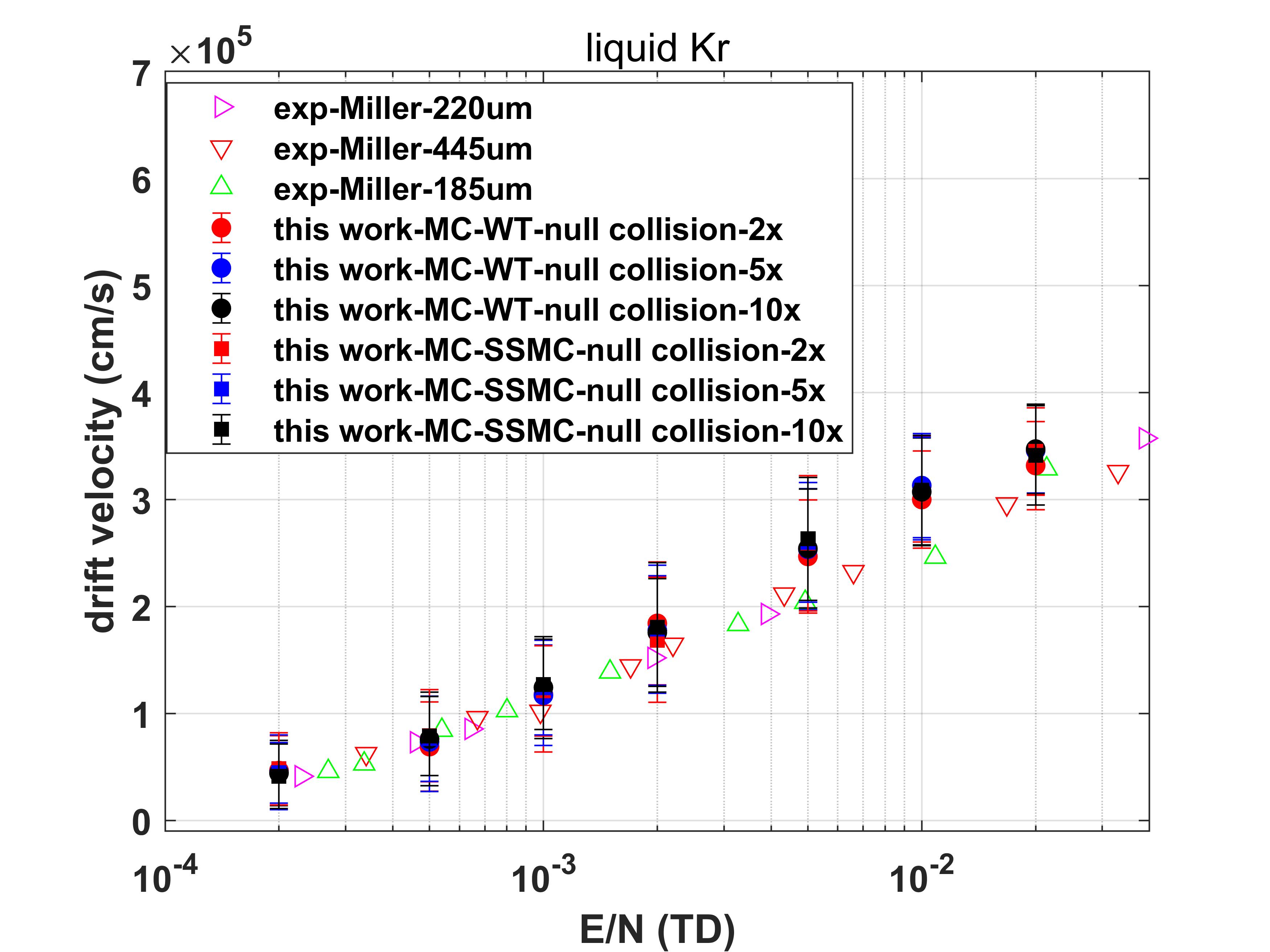}}
\label{fig_sub8.1}
  \hfil
\subfloat[the electron diffusion in LKr]{\includegraphics[width=0.45\textwidth]{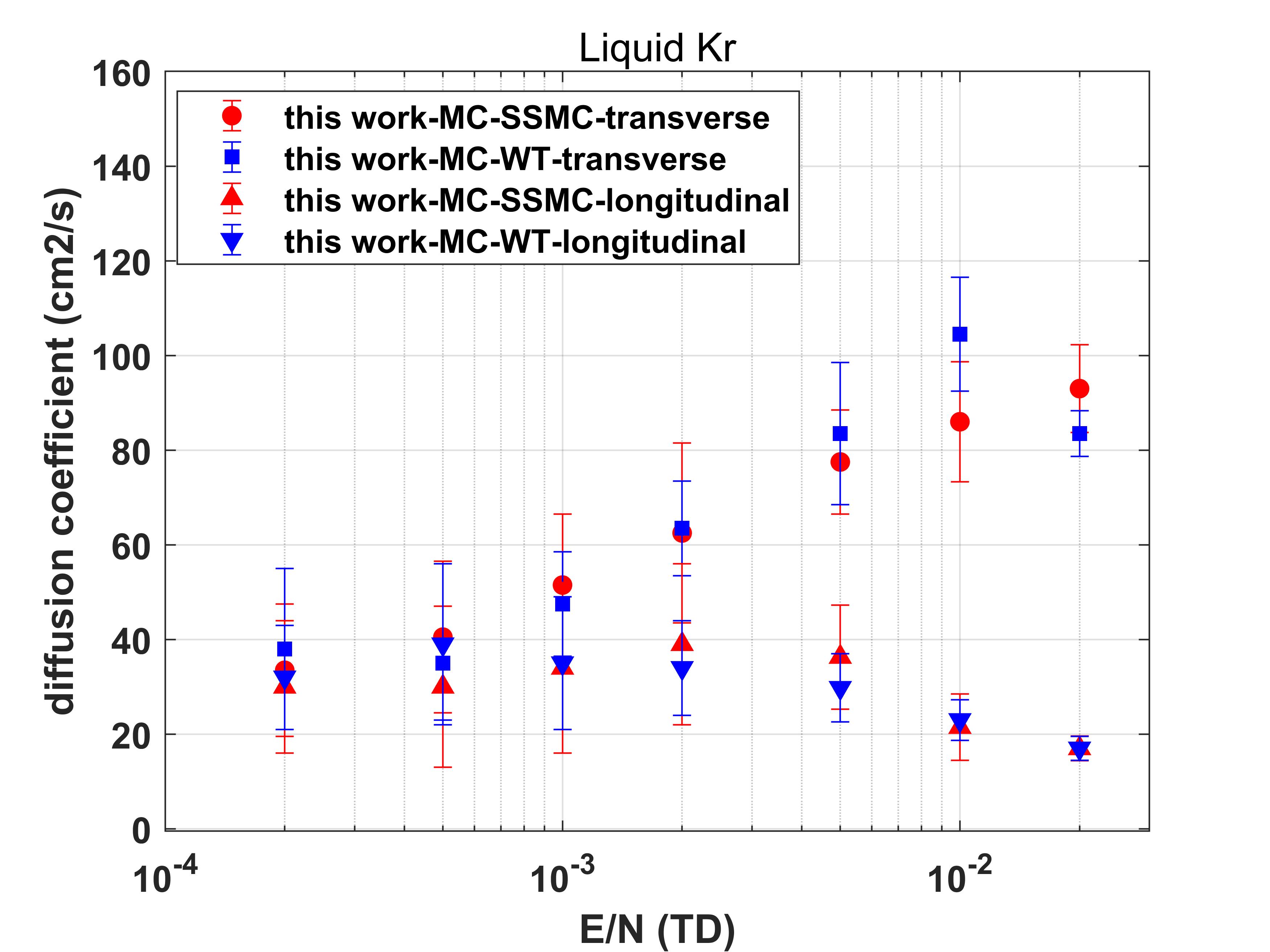}}
\label{fig_sub8.2}
 \caption{Electron swarm parameters in liquid krypton. "exp" represents experimental data, and "simulation" represents simulated data. (a) The electron drift velocity in liquid krypton. The experimental data are from Miller\cite{miller.PhysRev.166.871}. 220 um, 185 um and 445 cm are
the drift distances in Miller’s experiment. (b) The electron longitudinal and transverse diffusion in liquid krypton.}
   \label{fig:combined8}
\end{figure*}

\begin{figure*}[!ht]
\centering
\subfloat[drift velocity in LAr]{\includegraphics[width=0.32\textwidth]{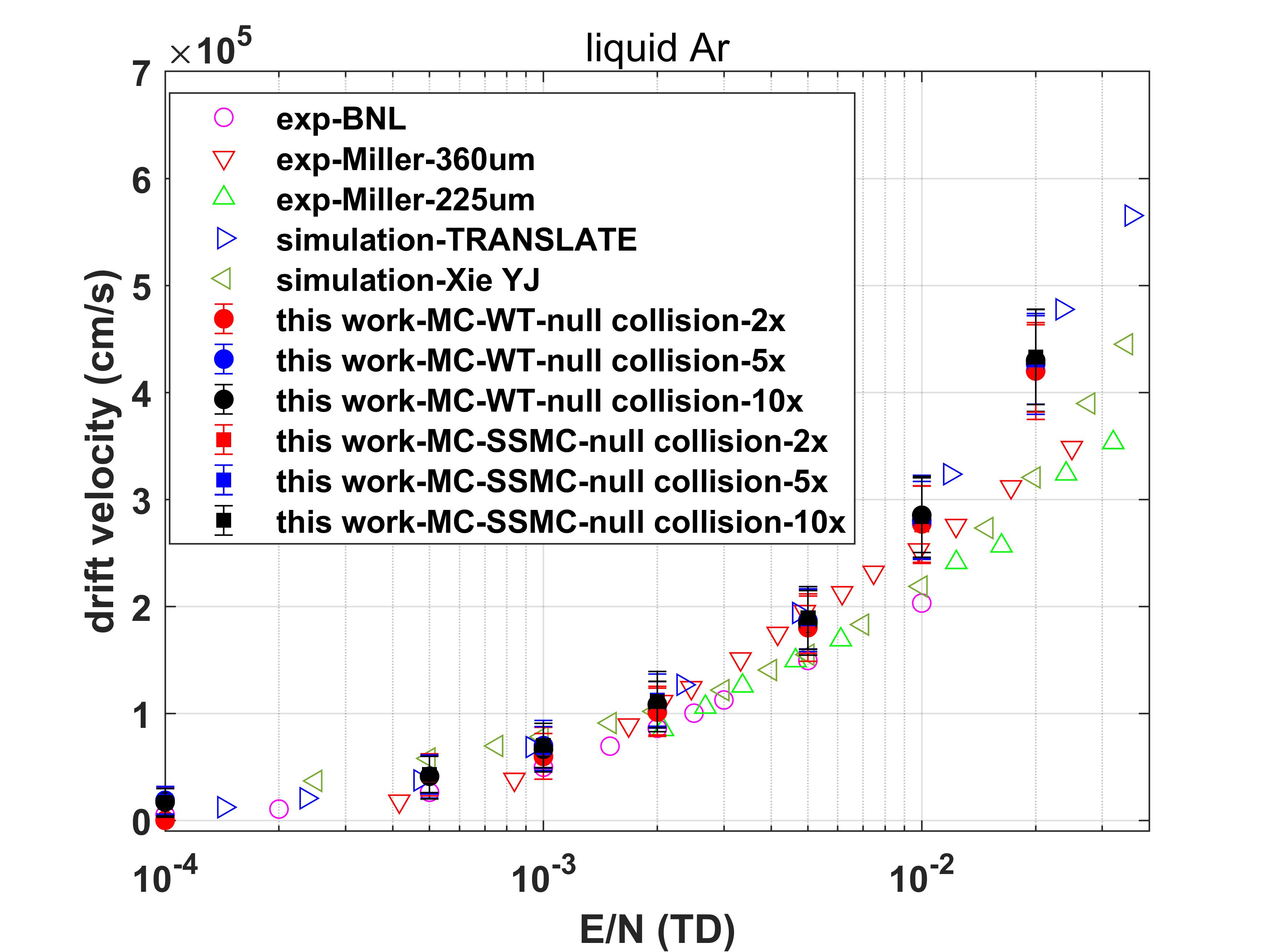}}
\label{fig_sub6.1}
  \hfil
\subfloat[longitudinal diffusion in LAr]{\includegraphics[width=0.32\textwidth]{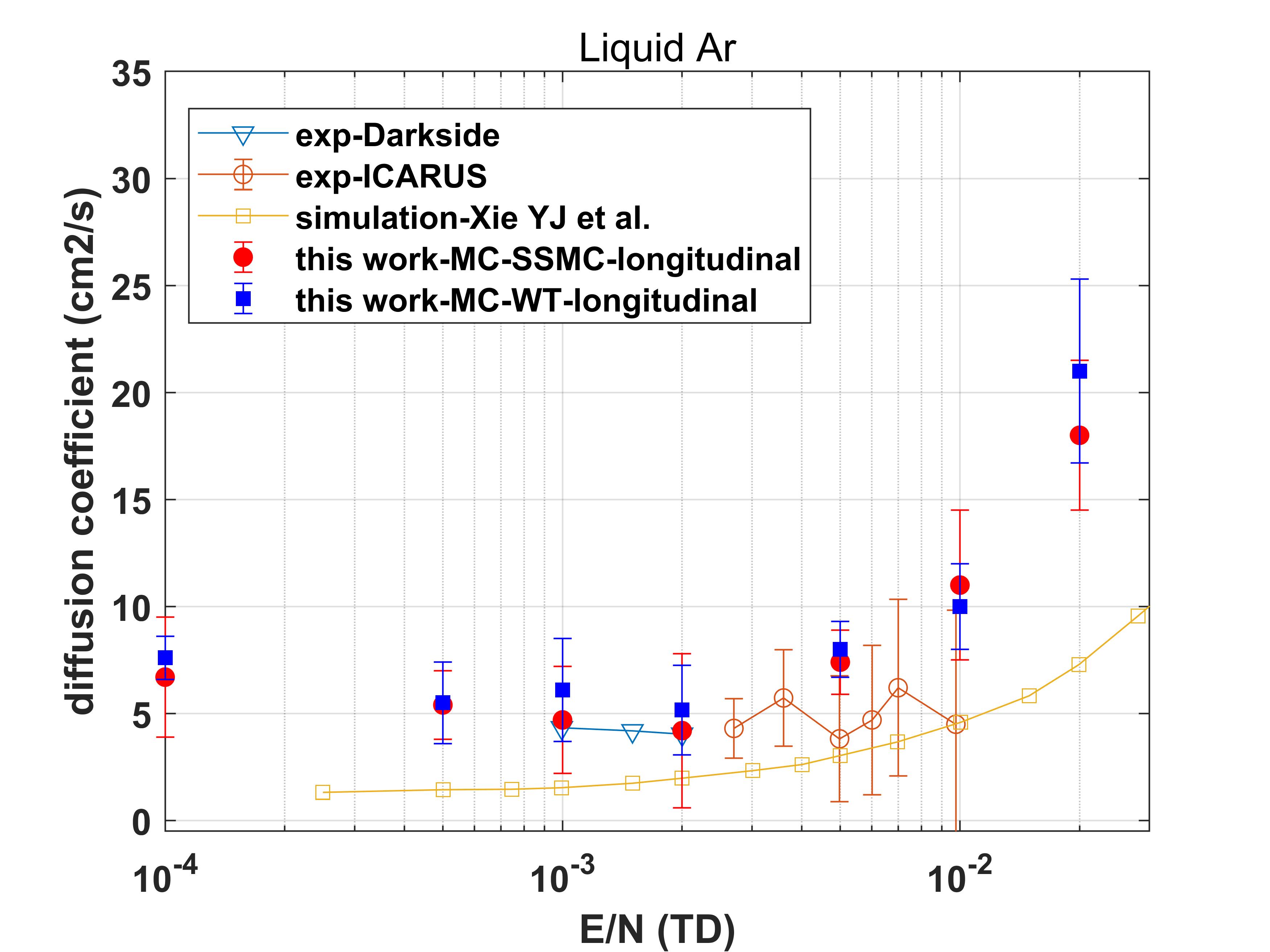}}
\label{fig_sub6.2}
  \hfil
\subfloat[transverse diffusion in LAr]{\includegraphics[width=0.32\textwidth]{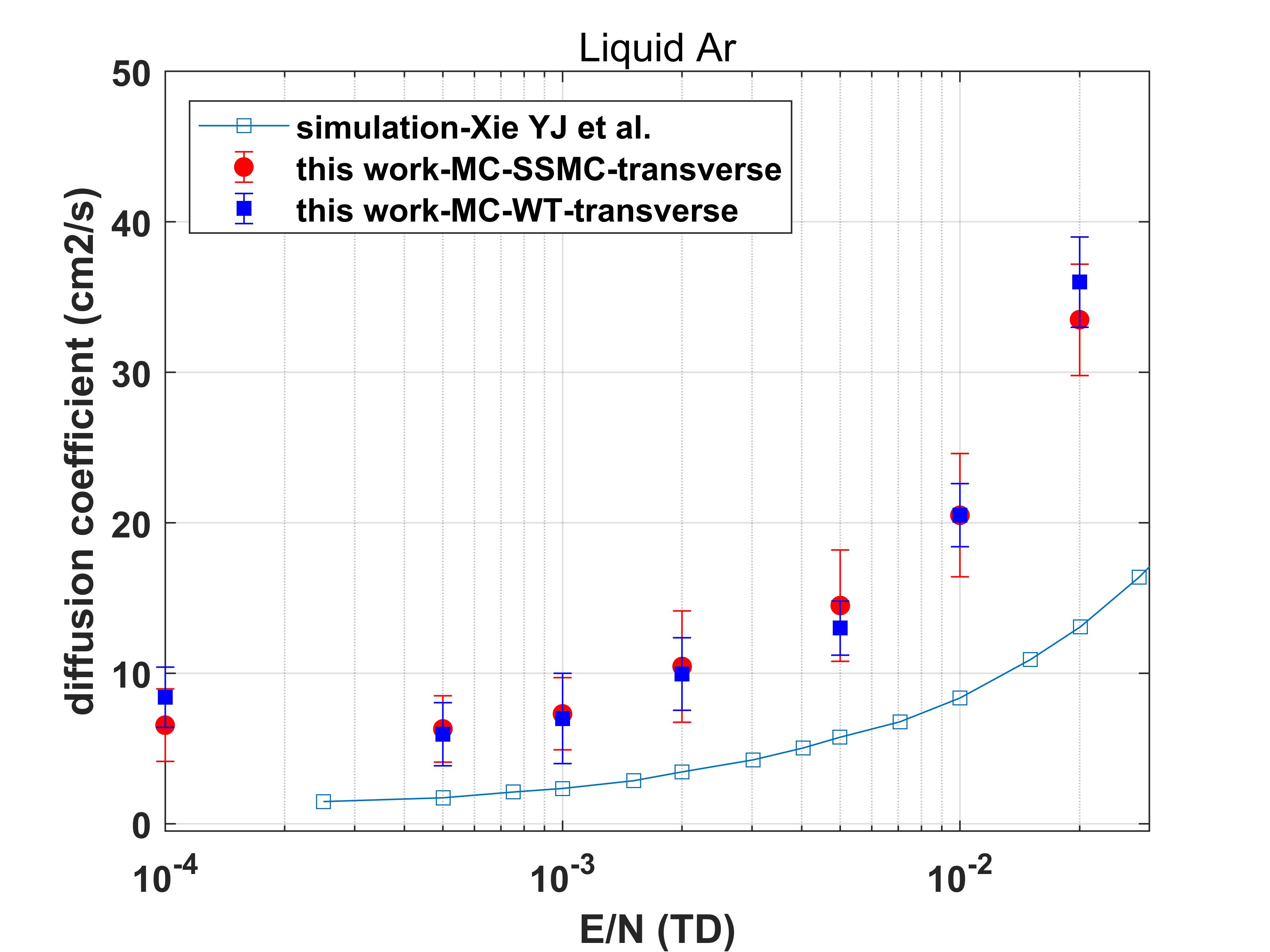}}
\label{fig_sub6.3}
 \caption{Electron swarm parameters in liquid Argon. "exp" represents experimental data, and "simulation" represents simulated data. (a) The electron drift velocity in liquid Argon. The experimental data are from BNL\cite{LI2016160}\cite{bnl}, Miller\cite{miller.PhysRev.166.871}. 360 um and 225 cm are the drift distances in Miller's experiment. The simulated data are from TRANSLATE\cite{beever2024translate} and Xie Yijun et al\cite{xie2024development}. (b) The electron longitudinal diffusion in liquid Argon. The experimental and MC data are from Darkside-50\cite{darkside50.AGNES201823}, ICARUS\cite{ICARUS.Torti_2017} and Xie YJ\cite{xie2024development}. (c) The electron transverse diffusion in liquid Argon.}
   \label{fig:combined6}
\end{figure*}

\begin{figure*}[!ht]
\centering
\subfloat[drift velocity in LXe]{\includegraphics[width=0.32\textwidth]{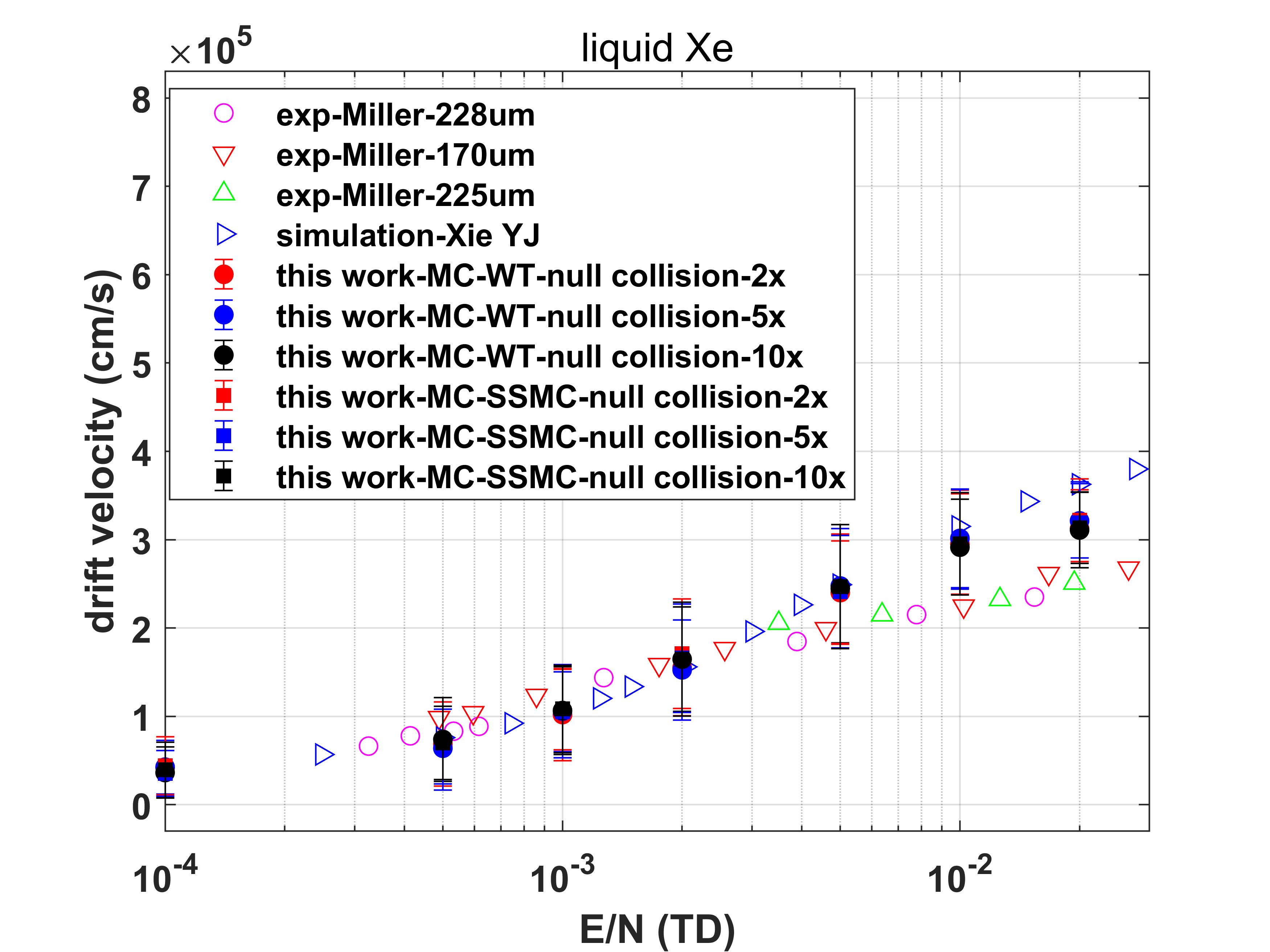}}
\label{fig_sub7.1}
  \hfil
\subfloat[longitudinal diffusion in LXe]{\includegraphics[width=0.32\textwidth]{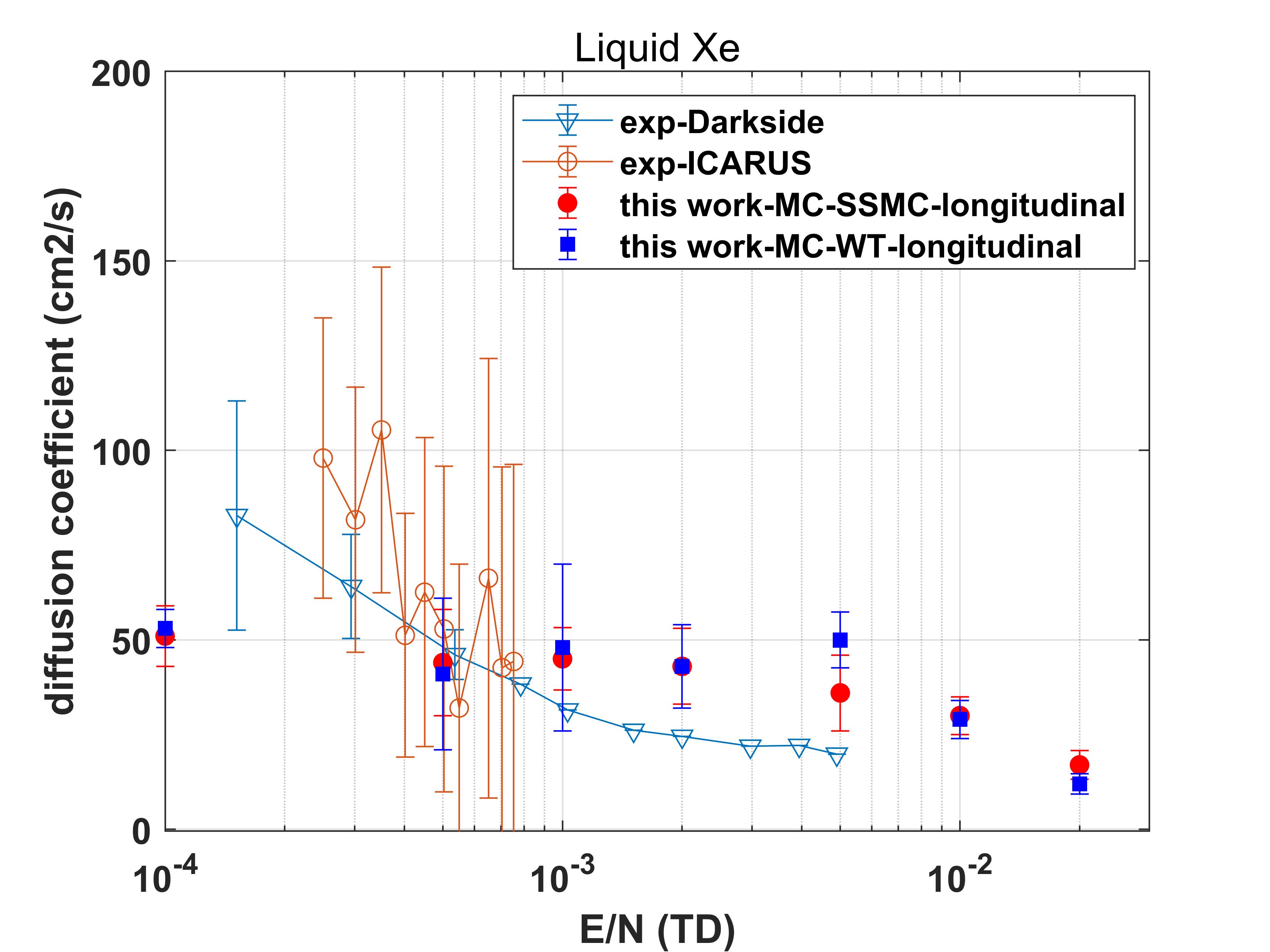}}
\label{fig_sub7.2}
  \hfil
\subfloat[transverse diffusion in LXe]{\includegraphics[width=0.32\textwidth]{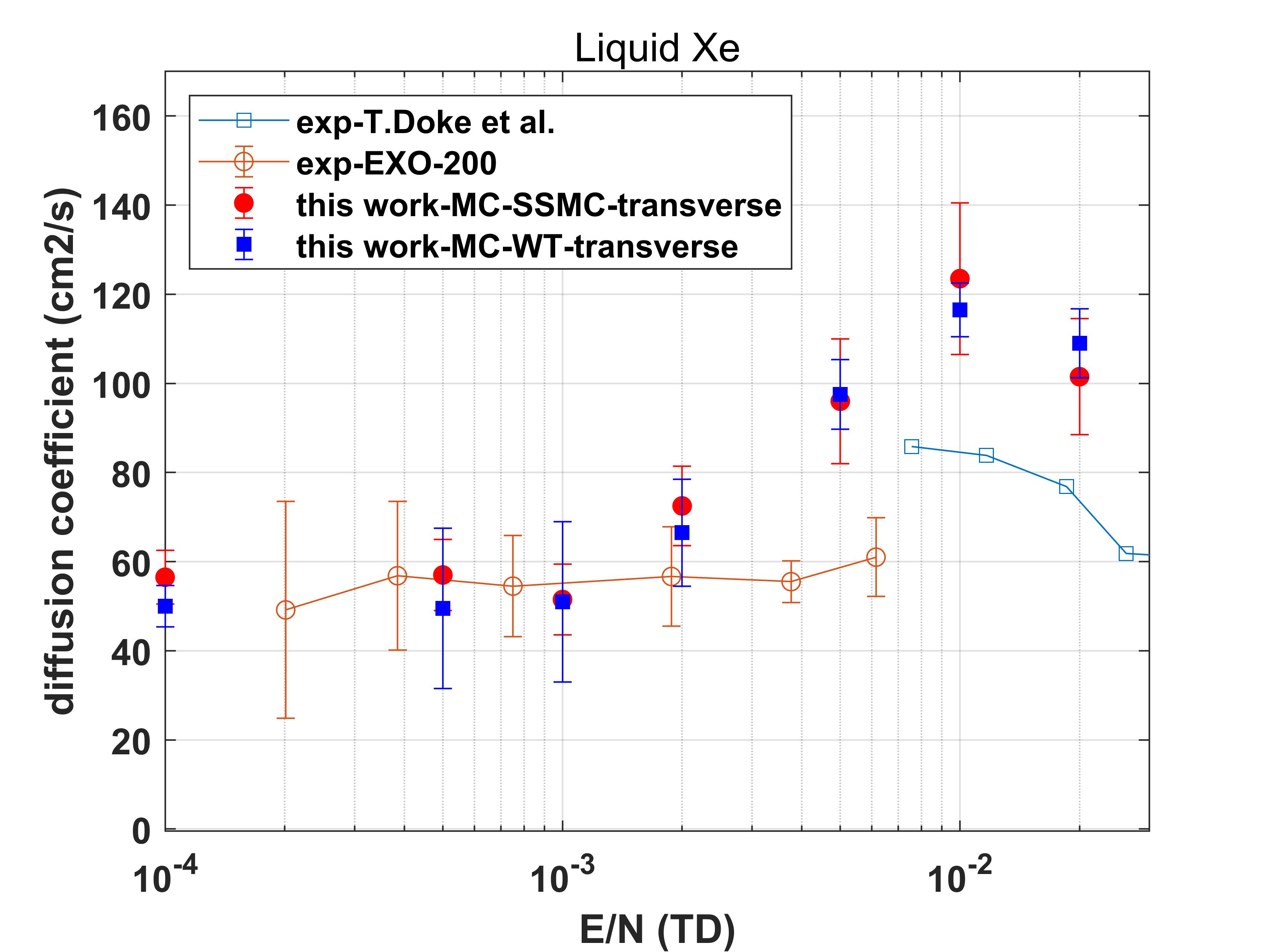}}
\label{fig_sub7.3}

 \caption{Electron swarm parameters in liquid xenon. The "exp" represents experimental data, and the "simulation" represents simulated data. (a) The electron drift velocity in liquid xenon. The experimental data are from Miller\cite{miller.PhysRev.166.871}. The simulated data are from Xie Yijun et al\cite{xie2024development}. (b) The electron longitudinal diffusion in liquid xenon. The experimental data are from Xenoscope\cite{Xenoscope.Baudis2023ElectronTransport} and Hogenbirk\cite{Hogenbirk_2018}. (c) The electron transverse diffusion in liquid xenon. The experimental data are from EXO-200\cite{EXO-200.PhysRevC.95.025502}, T.Doke\cite{DOKE198287}. }
   \label{fig:combined7}
\end{figure*}

In the liquid phase, the excitation and ionization cross sections are consistent with those in the gas phase. The elastic scattering cross section in the low-energy region turns to the energy transfer cross section and the momentum transfer cross section. This is illustrated for liquid argon\cite{4156740}, krypton\cite{4156740} and xenon\cite{4156740} in Fig.\ref{cross section in liquid}. The results of the simulation for the electron drift velocity (left panel) and diffusion coefficient (right panel) of liquid argon, krypton, and xenon are shown in Fig.\ref{fig:combined8}-\ref{fig:combined7}. 

\indent The effect of altering the collision frequency is not as apparent in the liquid phase as it is in the gas phase. This difference arises from the significantly higher atom density utilized in the liquid simulation, which reduces the mean free path by approximately two orders of magnitude. Consequently, the collision frequency is also about two orders of magnitude higher in the liquid phase. Thus, varying the collision frequency by a factor of 2x, 5x, and 10x, the effect was not as pronounced as in the gas phase. Therefore, the diffusion coefficients are shown in Fig.14-16 with 2x frequency.  

\indent The simulated drift velocity in the WT and SSMC models exhibits a high level of consistency as expected. In regions where the electric field exceeds 1000 V/cm, the simulated drift velocity slightly surpasses the experimental results, but overall remains within the margin of error.

\indent The reason for discrepancy between simulation results and the experimental data is as follows: One potential source of this discrepancy is that our simulation does not currently account for the electron recombination process, which could influence the diffusion dynamics. Additionally, the simulation assumes a pure environment, whereas the experimental setup may contain trace impurities that are not present in the simulation. These impurities could alter the electron transport properties and contribute to the deviation. 
\indent Regarding the diffusion coefficient, the experiments for the diffusion of noble liquids pose challenges with a large error bar.  
Upon comparing the simulation and experimental results, it can be seen that the diffusion coefficient shows the same trend and is of a comparable magnitude. For the diffusion coefficient, the error bars reflect the linear fitting error of the position sigma squared $\sigma^2$ versus time $t$. The validation of the drift velocity and diffusion coefficient underscores the reliability and accuracy of the simulation tool for noble liquids.

\subsection{Limitation}

\indent This simulation tool also has some limitations. It does not take into account the effect of electron recombination which in the electric field from 10 to 2000 V/cm has little effect on the simulation. Additionally, simulation for gas or liquid mixtures such as $Ar+CO_2$ for practical applications is not equipped. For mixtures, the attachment effects need to be considered, and this would be further improved in the subsequent work.

\indent Furthermore, this study does not include simulations for liquid neon and liquid helium. In these substances, electrons turn to a localized state known as "eBubble"\cite{PhysRevLett.28.1504}\cite{PhysRevLett.22.20}, the phenomenon arising from the Pauli exclusion principle. Notably, eBubble has a significantly lower drift velocity compared to free electrons\cite{SAKAI1992139}. Currently, there is no suitable Monte Carlo model available for simulating electron transport in liquid neon and helium. Therefore, we will focus on exploring the simulation of eBubble to address this limitation in the future investigations.

\section{Conclusion}
\label{section 4}
We explores a comprehensive electron transport Monte Carlo simulation tool designed for noble gases and liquids, including gaseous helium, neon, argon, krypton, xenon, and liquid argon, krypton, xenon. The simulation tool is developed using MATLAB and C++ to investigate electron behavior. Particularly in the liquid phase, the WT and SSMC models are used for the electron elastic scattering simulation. Validation of the simulation tool is achieved through the electron swarm parameters: drift velocity and diffusion coefficient for electric fields ranging from 10 to 2000 V/cm. The simulation tool provides valuable support for gas or liquid detector design applications.

\section{Acknowledgements}
This work is supported in part by National Key Research and Development Program of China (Grant No.2023YFA1607204), the real-time monitoring technology of disease treatment based on micro-flash therapy instrument (No.74140-71020002), and Development and application of micro-flash therapy instrument in collaboration with near-infrared light energy conversion for tumor treatment (No.74140-71020005). This work is also supported by Guangdong Provincial Key Laboratory of Advanced Particle Detection Technology. Additionally, this work is supported by the "Sun Yat-Sen University Gamma Photon Collider and Comprehensive Beam Facility (Phase I) Project," with the project code 2403-000000-05-03-714165.

\indent 
The authors would like to express sincere gratitude to Prof. Wang Yi and Prof. Qi Huirong for their valuable suggestions.


\end{document}